\documentclass[twoside,11pt]{article}

\usepackage{jmlr2e}

\usepackage{pdflscape}
\usepackage{multirow}
\usepackage{graphicx}
\usepackage{amsmath}
\usepackage{amssymb}
\usepackage{amstext}
\usepackage{subfigure}
\usepackage{color}
\usepackage{mathrsfs} 
\usepackage{natbib} 
\usepackage{booktabs}
\usepackage{array}
\usepackage{bm}

\newcolumntype{P}[1]{>{\centering\arraybackslash}p{#1}}

\usepackage{xspace} 
\newcommand{\rmax}{R-\textsc{max}\xspace}
\newcommand{\rmaxsharp}{R-\textsc{max}\#\xspace}
\newcommand{\bprnew}{BPR+\xspace} 
\newcommand{\qlearning}{\emph{Q}-learning\xspace} 

\DeclareMathOperator*{\argmax}{arg\,max}

\usepackage{mathtools}

\usepackage[ruled,vlined]{algorithm2e}
\usepackage{changes}

\ShortHeadings{A Survey of Learning in Multiagent Environments}{Hernandez-Leal, Kaisers, Baarslag and Munoz de Cote}
\firstpageno{1}

\begin{document}

\title{A Survey of Learning in Multiagent Environments:\\ Dealing with Non-Stationarity}

\author{\name Pablo Hernandez-Leal  \email Pablo.Hernandez@cwi.nl \\
 \name Michael Kaisers  \email Michael.Kaisers@cwi.nl \\
 \name Tim Baarslag  \email Tim.Baarslag@cwi.nl \\
 		\addr Intelligent and Autonomous Systems Group \\
        \addr Centrum Wiskunde \& Informatica \\
 			 Amsterdam, The Netherlands
         \AND  
        \name Enrique Munoz de Cote \email jemc@inaoep.mx \\       \addr Instituto Nacional de Astrof\'isica, \'Optica y Electr\'onica, Puebla, M\'exico \\
       %Luis Enrique Erro 1, Sta. Mar\'ia Tonantzintla, Puebla, M\'exico\\
      \addr PROWLER.io Ltd., Cambridge, United Kingdom
      }

\editor{}

\maketitle

\begin{abstract}%   <- trailing '%' for backward compatibility of .sty file
The key challenge in multiagent learning is learning a best response to the behaviour of other agents, which may be non-stationary: if the other agents adapt their strategy as well, the learning target moves. Disparate streams of research have approached non-stationarity from several angles, which make a variety of implicit assumptions that make it hard to keep an overview of the state of the art and to validate the innovation and significance of new works.
This survey presents a coherent overview of work that addresses opponent-induced non-stationarity with tools from game theory, reinforcement learning and multi-armed bandits. Further, we reflect on the principle approaches how algorithms model and cope with this non-stationarity, arriving at a new framework and five categories (in increasing order of sophistication): \emph{ignore, forget, respond to target models, learn models,} and \emph{theory of mind}. A wide range of state-of-the-art algorithms is classified into a taxonomy, using these categories and key characteristics of the environment (e.g., observability) and adaptation behaviour of the opponents (e.g., smooth, abrupt). To clarify even further we present illustrative variations of one domain, contrasting the strengths and limitations of each category. Finally, we discuss in which environments the different approaches yield most merit, and point to promising avenues of future research.
\end{abstract}


% !TEX root = survey.tex

\section{Introduction}

% FOR SURVEY PAPERS ONLY:
% - IS A SURVEY IN THIS AREA TIMELY?    [IS THE AREA OF INTEREST TO 
%   THE AUTONOMOUS AGENTS AND MULTIAGENT SYSTEMS COMMUNITY AND IS THERE NO RECENT AND COMPREHENSIVE 
%   REVIEW ALREADY AVAILABLE?] 
% - IS THE COVERAGE OF THE AREA BALANCED, COMPLETE AND UP-TO-DATE? 
% - DOES THE SURVEY PROVIDE A SUITABLE FRAMEWORK FOR UNDERSTANDING THE AREA? 
%
%\revise{- throughout the paper: you write both state-of-the-art and state of the art. choose one and use consistently - throughout the paper: often written stationary where it should be stationarITY }

%\revise{Tim: I think it is fine like it is now;  we use ``state-of-the-art'' as an adjective and ``state of the art'' as a noun. Pablo: Agree}

\noindent
There are many successful applications of multiagent systems (MAS) in the real world. Examples are ubiquitous in energy applications, for example, to implement a network to distribute electricity~\citep{Pipattanasomporn:2009vq} or to coordinate the charging of electric vehicles~\citep{Valogianni:2015tq}, in security,  to patrol the Los Angeles airport~\citep{Pita:2009uw} and in disaster management to assign a set of resources to tasks~\citep{Ramchurn:2010uu}. Multiagent systems include a set of autonomous entities (agents) that share a common environment and where each agent can independently perceive the environment, act according to its individual objectives and as a consequence, modify the environment. 

How the environment changes as a consequence of an agent exerting an action is known as the \textit{environment dynamics}. In order to act optimally with respect to its objectives these dynamics need to either be known (a priori) by the agent or otherwise be learned by experience, i.e., by interacting many times with the environment. Once the environment dynamics have been learned, the agent can then \textit{adapt} its behaviour and act according to its target objective. We know a lot about the single agent case, where only one agent is learning and adapting its behaviour, but most of the results break apart when two or more agents share an environment and they all learn and adapt their behaviour concurrently. 

The problem with this concurrency is that the action executed by one agent affects the goals and objectives of the rest, and vice-versa. To tackle this, each agent will need to account for how the other agents are behaving and  adapt according to the \textit{joint behaviour}. Needless to say, this joint behaviour needs to be learned by each agent, and due to the fact that all agents are performing the same operations of learning and adapting concurrently, the joint behaviour ---and therefore the environment---  is perceived by each agent as \textit{non-stationary}. This non-stationarity (sometimes referred to as the moving target problem, see~\citealp{Tuyls:2012up}) sets multiagent learning apart from single-agent learning, for which it suffices to converge to a fixed optimal strategy.
%%%TODO: add http://web.mit.edu/jnt/www/Papers/J052-94-jnt-q.pdf
%multiple agents are present converge guaranteees of RL no longer hold, agents are in a non-stationary environment
%

Most learning algorithms to date are not well suited to deal with non-stationary environments,\footnote{Environments, in which all counterpart agents are perceived as part of the environment.} and usually, such non-stationarity is caused by changes in the behaviour of the participating agents. For example, a charging vehicle in the smart grid might change its behavioural pattern~\citep{Marinescu:2015wt}; robot soccer teams may change between pre-defined behaviours depending on the situation~\citep{MacAlpine:2012vn}; and attackers change their behaviours to keep security guards guessing in domains involving frequent adversary interactions, such as wildlife and fishery protection~\citep{fang2015defender}.
%and poker playing~\citep{Bard:2013ta}. 
%Ignoring or misunderstanding the implicit assumptions made by different algorithms can lead to serious consequences. For example, in power systems, deterministic responses intended to dampen the system may actually result in self-reinforcing oscillations, resulting in instability of the system with grave consequences \citep{Ernst:2004ih}. 
%TODO: 
%% CONTINUOUS ADAPTATION VIA META-LEARNING IN NONSTATIONARY AND COMPETITIVE ENVIRONMENTS

Previous works in reinforcement learning (RL), MAS and multi-armed bandits (to name a few) have all acknowledged the fact that specialized targeted work is needed that explicitly addresses non-stationary environments~\citep{Sutton:2007ha,Panait:2005wj,Garivier:2011br,Matignon:2012bj,Lakkaraju:2017to}.
Against this background, this survey fills this gap with an extensive analysis of the state of the art. 
Previous surveys have proposed different ways to categorise MAS algorithms~\citep{Panait:2005wj,Shoham:2007vw,Busoniu:2010ft}, others have divided them by the type of learning \citep{Tuyls:2012up,Bloembergen:2015ei} and another group have proposed properties that MAS algorithms should have \citep{Bowling:2002vva,Powers:2007gq,Crandall:2011dt}. In contrast, we propose another view, which has been mostly neglected, focused on how algorithms deal with non-stationarity, providing an illustrative categorization with increasing order of sophistication where each algorithm is analysed along with related characteristics (observability and opponent adaptation).

The questions addressed by the surveyed algorithms are illustrated by the following simple scenario comprising two agents:
\begin{description}
	\item[Predator.] The agent under our control.
	\item[Prey.] The opponent agent.\footnote{In this work we use the word ``opponent" when referring to another agent in the environment irrespective of the domain, and irrespective of its adversarial or cooperative nature.} 
\end{description}
Both agents engage in repeated rounds of interactions (possibly infinite), there is no communication and the rewards received depend on the joint action. The prey has several (possibly infinite) strategies at its disposal (ways to select its actions) and it can change from one to another during the interaction. In this context we can raise several questions:

\begin{itemize}
\item Should the predator assume the prey will behave in a certain way (e.g., minimizing the predator's reward, enacting their part of the Nash Equilibrium or playing as a teammate)?
\item Should the predator learn to optimise against a single opponent strategy or should it generalise by learning a more robust strategy against a class of possible strategies? 
\item Should the predator assume the prey is modelling the predator's strategy?
\item Should the predator assume the prey will use a stationary strategy? If not, will the prey change its behaviour slowly or drastically?
\end{itemize}
Different research communities make different assumptions that give rise to distinct answers to these questions. While there is some awareness within each community of the work outside that community, it remains a challenge to keep up to date with the recent literature due to this fragmentation, which impedes AI research in its entirety~\citep{Eaton:2016kz}.

For example, many game theory algorithms focus on finding equilibria in self-play. Multi-armed bandits either assume a stochastic or adversarial setting and try to optimize against that behaviour. Some basic approaches of reinforcement learning ignore other agents and optimise a policy assuming a stationary environment, essentially treating non-stationary aspects like stochastic fluctuations. Other approaches learn a model of the other agents to predict their actions to remove the non-stationary behaviour. Finally, algorithms from behavioural game theory and planning have proposed recursive modelling approaches that assume opponents are capable of performing strategic reasoning and modelling of the rest of the agents. 

%\comment{JAIR: address the following issues: (1) analyze a significant body of AI research and make it more accessible to a broader audience; (2) position existing research results in a broader context and explain their impact; (3) bring together previously unconnected lines of research in a way that fosters new research directions in these areas; (4) identify deficiencies or gaps in current knowledge that need to be addressed in future research.}

In this context, the main contributions of this survey are the following: 
\begin{itemize}
\item Provide a coherent view of how state-of-the-art algorithms in reinforcement learning, multi-armed bandits and game theory tackle the planning problem of long-term sum of expected rewards in non-stationary environments. 
\item Propose a new framework for multiagent systems (see Section~\ref{sec:newFramework}). This framework allows to describe a categorisation with increasing order of sophistication with respect to how non-stationarity is handled, arriving at five categories: ignore, forget, respond to target opponents, learn opponent models and theory of mind (see Section~\ref{sec:categories}).
\item Describe the fundamental algorithms of each category using an illustrative example highlighting their strengths and limitations (see Section~\ref{sec:example}).
\item Categorise most significant learning algorithms while also describing their main characteristics with respect to the environment and opponent assumptions (see Section~\ref{sec:learningnonstationary}). 
\item Provide a structured set of open questions with promising avenues of future research in multiagent learning (see Section~\ref{sec:openquestions}).
\end{itemize}
With its tailored scope, this survey aims to establish a structure to think clearly about all the assumptions, characteristics and concepts related to the challenge of addressing non-stationarity in multiagent learning.

\subsection{Related work and demarcation}
% Since you have a related work section covering prior surveys, please also be aware of this one:
%http://www.cs.utexas.edu/~pstone/Papers/bib2html/b2hd-MASsurvey.html
%
%Btw. Peter and I gave a tutorial on multiagent learning at IJCAI'17 and 
%the slides are available online 
%(http://www.cs.utexas.edu/~larg/ijcai17_tutorial). Maybe worth mentioning somewhere in your survey?

%Another survey 
%Learning in Nonstationary Environments: A Survey
%http://citeseerx.ist.psu.edu/viewdoc/download?doi=10.1.1.728.3478&rep=rep1&type=pdf
% In supervised learning, nonstationarity is often reffered as concept drift.
%
%

Multiagent learning has received a lot of attention in the past years and some previous surveys have emerged with different motivations and outcomes. \cite{Shoham:2007vw} presented a general survey of multiagent learning providing some interesting foundational questions and identifying five different agendas in this research community. \cite{Tuyls:2012up} presented a bird's eye view about the AI problem of multiagent learning, identifying the milestones achieved by the community and mentioning the open challenges at the time.\footnote{We reflect on the relation between those challenges and the promising avenues of future research in Section~\ref{sec:openquestions}.} \cite{Panait:2005wj} presented an extensive analysis of cooperative multiagent learning algorithms, dividing them into two categories: single learner in a multiagent problem (team learning) and multiple learners (concurrent learning). \cite{Matignon:2012bj} focused on the evaluation of independent RL algorithms on cooperative stochastic games. \cite{Busoniu:2010ft} presented a thorough survey on multiagent RL where they identified a taxonomy and several properties for algorithms in multiagent reinforcement learning (MARL). \cite{Crandall:2011dt} assessed the state of the art in two-player repeated games with respect to three properties: security, cooperation and compromise, which they propose as important to act in a variety of different games. \cite{Muller:2014uo} presented an application-oriented survey, highlighting applications that use or are based on MAS. \cite{Weiss:2013ue} edited a book about multiagent systems; in particular there is a chapter dedicated to multiagent learning where they present state-of-the-art algorithms dividing them into joint action, gradient, Nash and other learners (\citealp[see][chap. 10]{Weiss:2013ue}). A recent survey analysed methods from evolutionary game theory and its relation with multiagent learning~\citep{Bloembergen:2015ei}. %\revise{\cite{da2019survey} presented a survey on the subject of knowledge reuse in multiagent RL.}
%\reviseq{Not mentioned by the reviewers but we might to highlight that do not cover muultiagent deep RL}
%\revise{
Finally, the recent area of multiagent deep reinforcement learning gained a lot of interest with two recent surveys~\citep{nguyen2018deep,hernandez2018multiagent}. %} 
None of these survey articles provide an explicit treatment of the non-stationarity approaches taken in various algorithms.

Our survey exceeds previous work in scope of different domains and coverage measured by number of algorithms, and fills the gap of reflecting on non-stationarity.
In contrast to previous works, we provide a detailed analysis of algorithms from multi-armed bandits (for stochastic and adversarial environments), single agent RL (model-based and model-free approaches), multiagent RL and game theory (mainly for repeated and stochastic games) in both competitive and cooperative scenarios. We provide a full taxonomy of how algorithms cope with non-stationarity, and describe opponent and environment characteristics.

This survey does not cover work related to learning in dynamic environments that do not have other active autonomous and automated agents, such as recommender systems of news articles~\citep{Liu:2010vj} or online supervised learning scenarios (see Section~\ref{sec:relatedModels}). 
 
\subsection{How to read this survey}

Since different audiences are expected to read this survey, this section provides forward references to key insights and sections for different target groups:
\begin{itemize}
\item For \textbf{researchers seeking an introduction} to multiagent learning we propose to follow the current structure of the paper sequentially, progressing through each section in order.

\item For \textbf{experienced researchers} we recommend starting with the \textcolor{black}{new framework proposed in Section \ref{sec:newFramework}}, followed by the high level vision of the categorisation of algorithms depicted in Figure~\ref{fig:categories} in Section~\ref{sec:example}; to be followed by the extensive categorization in Section~\ref{sec:learningnonstationary}; in particular given in Table~\ref{tab:stateoftheart} and Figure~\ref{fig:relatedWork}. 

\item We encourage \textbf{researchers seeking guidance on promising research directions} to consult the discussion in Section~\ref{sec:discussion}, in particular to find common types of results in Section~\ref{sec:theoreticalresults} and interesting open problems in Section~\ref{sec:openquestions}.% \comment{More recommended sections? same for previous bullet}
\end{itemize}
Finally, we encourage all readers to position their future work in this framework, as delineated in Section~\ref{sec:learningMAS}, for ease of reference and navigation of related (future) work.

\subsection{Paper overview}

This paper aims to provide a general overview of how different algorithms cope with the problem of learning in multiagent systems where it is necessary to deal with non-stationary behaviour. In Section~\ref{sec:preliminaires}, we review formal models used in this context %\revise{and we highlight in which context they are appropriate for}
; in particular we review multi-armed bandits, reinforcement learning and game theory. \textcolor{black}{Section~\ref{sec:learningMAS} describes the main challenge of non-stationarity in multiagent systems together with a new framework that naturally models its key elements, and lastly %describes a taxonomy of the environment characteristics (observability), list assumptions about the opponent adaptation and 
presents the proposed categorization of how algorithms deal with non-stationarity}. Section~\ref{sec:example} illustrate the categories using a simple scenario. Section~\ref{sec:learningnonstationary} presents an extensive list of works of multi-armed bandits, RL and game theory categorised by the taxonomy proposed in this survey. Section~\ref{sec:discussion} provides a discussion about the strengths and limitations of each category, describes the common experimental settings, presents a summary of the theoretical results and pinpoints interesting open problems highlighting promising lines of future research. Finally, Section~\ref{sec:conclusions} summarizes the conclusions and contributions of this survey.

%% A Deep Policy Inference Q-Network for Multi-Agent Systems

% !TEX root = survey.tex

\section{Formal approaches from different domains that model non-stationarity}
\label{sec:preliminaires}

This section describes the formal models used in multi-armed bandits, reinforcement learning, and game theory, and contrasts how they capture non-stationarity. Each domain makes different assumptions about a priori information about the interaction, as well as about \emph{online} observability of the environment and opponents during the interaction. This discrimination forms the basis of the environment characteristics in the next section. In line with available information, the solution concept for finding good behaviour may be characterized correspondingly by more a priori or more online reasoning, exhibiting characteristic behaviour and the ability to cope with certain types of non-stationarity in opponent behaviour. In order to present our behavioural categorization, we here present a synopsis of the different approaches from the literature.

Note that different areas provide different terminology. Therefore, we will use the terms player and agent interchangeably; similarly for reward and payoff; and for rounds and steps. Finally, we will refer to other agents in the environment as opponents irrespective of the domain's or agent's cooperative or adversarial nature.

\subsection{Multi-armed bandits}

The simplest possible reinforcement-learning problem is known as the multi-armed bandit problem \citep{Robbins:1985um}: the agent is in a room with multiple gambling machines (called ``one-armed bandits''). At each time-step the agent pulls the arm of one of the machines and receives a reward. The agent is permitted a fixed number of pulls. The agent's purpose is to maximise its total reward over a sequence of trials. Usually each arm is assumed to have a different distribution of rewards, therefore, the goal is to find the arm with the best expected return as early as possible, and then to keep gambling using that arm.

A $K-$armed (stochastic) bandit can be formalised as a set of real distributions $\mathcal{B}=\{R_1,R_2,\dots, R_k\}$, with the set of arms $I = \{1, \ldots, K\}$, such that each arm yields a stochastic reward $r_i$ following the distribution $R_i$. Let $\mu_1,\mu_2,\dots,\mu_k$ be the mean values associated with these reward distributions. A policy, or allocation strategy, is an algorithm that chooses the next machine to play based on the sequence of past plays and obtained rewards. The policy selects one arm at each round and observes the reward, this process is repeated for $T$ rounds. This problem illustrates the fundamental trade-off between exploration and exploitation: should the agent choose the arm with the highest average reward observed so far (exploit), or should it choose another one for which it has less information, so as to find out if it in fact exceeds the first one (explore)? 

The \emph{regret} $\Delta_R$ is a common measure used to evaluate different algorithms in multi-armed bandits. The regret is the difference (necessarily a loss) between the chosen policy $\pi$ and the optimal policy $\pi^*$. In the multi-armed bandit setting the optimal policy would choose the arm $i^*$ with the highest expected reward at all times, i.e., $\forall t: \pi^*_t = i^*$, while $\pi_t = i(t)$ may vary over time. For an episode of $T$ steps,
the stochastic regret yields
$$  \Delta_R = \sum_{t=1}^T{ r_{i^*} } - \sum_{t=1}^T{ r_{i(t)}}.$$
% the expected regret yields
% $$  \Delta_R = T \mu_{i^*} - E \left[ \sum_{t=1}^T{ r_{i(t)}} \right].$$
% 
With this concept in mind, some approaches guarantee low regret under certain conditions.
These policies work by associating a quantity called \emph{upper confidence index} to each arm. This index relies on the sequence of rewards obtained so far from a given arm and is used by the policy as an estimate for the corresponding reward expectation~\citep{Auer:2002fd}. The UCB1 (Upper Confidence Bounds) algorithm achieves logarithmic regret assuming bounded rewards, without further constraints on the reward distributions~\citep{Auer:2002fd}. UCB uses the principle of \emph{optimism in the face of uncertainty} to select its actions, i.e., the algorithm selects arms by an optimistic estimate on the expected rewards of certain arms to balance exploration and exploitation. The UCB1 algorithm is extremely simple, it initially plays each arm once, and subsequently selects the arm~$i(t)$:
$$i(t) = \argmax_j \left( \bar r_j + \sqrt{\frac{2~\text{ln}~t}{n_j}} \right) $$
where $\bar r_j$ is the average reward obtained from arm $j$, $n_j$ is the number of times arm $j$ has been played, and $t$ is the total number of rounds so far.

\begin{algorithm}
\KwIn{$K$ arms, $T$ rounds ($T \ge K \ge 2$).}
\For{t=$1,\dots,T$}{
\nl Algorithm chooses one arm $i(t) \in \{1,\dots,K\}$\\
\nlset{2-a} \hspace{1cm} Adversary selects rewards $g_t=(g_{1,t},\dots,g_{K,t}) \in [0,1]^K$ \\
\nlset{2-s} \hspace{1cm} Stochastic environment produces reward $g_{i,t} \sim R_i$ (drawn independently)\\
\nlset{3} Receive reward $g_{i(t),t}$ (does not observe the other arms)\\
}
The goal is to minimise the regret, defined by :\\
\hspace{1cm} In the adversarial model, $\Delta_r= max_{j \in \{1,\dots, K \}} \sum_{t=1}^T g_{j,t} - \sum_{t=1}^T  g_{i(t),t} $\\
\hspace{1cm} In the stochastic model, $\Delta_r= \sum_{t=1}^T (max_{j \in \{1,\dots,K \}} \mu_{j}  - \mu_{i(t)} ) $\\
\caption{Multi-armed bandit: stochastic (s) or adversarial (a)~\citep{Bubeck:2012ue}}
\label{algo:mab}
\end{algorithm}

The stochastic bandit scenario is useful to model decision-making in stationary but stochastic settings. A direct extension of this setting is the \emph{adversarial model}, which assumes that rewards of each arm are controlled by an \emph{adversary}, i.e., the reward distribution associated with each arm at every round is fixed in advance by an adversary before the game starts~\citep{Auer:2002vx}; see Algorithm~\ref{algo:mab} that juxtaposes both scenarios. However, when relaxing the assumptions made in the problem definition even more by assuming \emph{online adaptive adversaries}, the standard definition of regret is no longer adequate (due to adaptivity of the adversary, the optimal action might change at different steps, see~\citealp{Arora:2012ta}). Because of that, different variations of the regret measure have been proposed~\citep{Arora:2012ta,Crandall:2014wx}. It is worth mentioning that there are further extensions to the bandit scenario~\citep{Pandey:2007eb,Beygelzimer:2011wp,TranThanh:2012uk}, which are beyond the scope of this survey. However, we refer the interested reader to the discussion in related work~\citep{Bubeck:2012gx}.

%A further notion of regret is the regret for the best \emph{expert} in a fixed set of experts (i.e., strategies that select actions) that are available to the player. The notion of expert is defined as a recommendation, in the form of a probability distribution over the arms, as to which bandit to play next. 

%In the context of adversarial bandits, the disappointment or policy regret \citep{Arora:2012ta} is defined to measure the success as: an expert algorithm should perform at least as well as it would have performed had it always followed its best expert, minus the assumption that agent i’s actions do not impact the adversarial future actions

%$$ \Delta_D = \sum_{t=1}^T u_i^t(\pi^+, \pi_{-i}^t(\pi^+)) - \sum_{t=1}^T{ \pi_t}$$
%
%where $u_i^t(\pi^+_i, \pi_{-i}^t(\pi^+))$ represents the agents $i$'s payoff using policy $\pi^+$, and the agent $-i$ using $\pi_{-i}^t(\pi^+)$, i.e., the policy used by agent $-i$ in time $t$ if agent $i$ had used $\pi^+$ up to round $t$. 

\subsection{Reinforcement Learning}

\begin{figure}[]
\centering
	\includegraphics[scale=0.45]{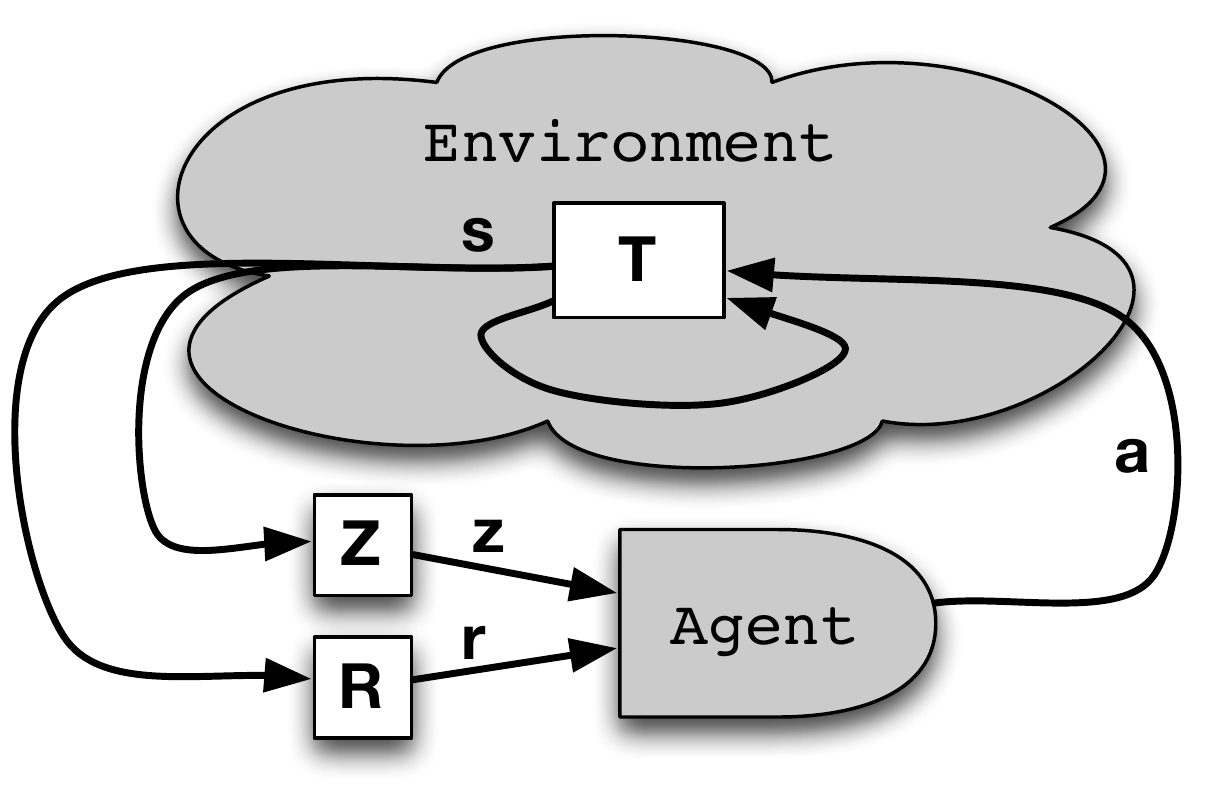}
	\caption{\small The agent interacts with the environment, performing an action $a$ that affects the state environment according to a function $T$, producing the state $s$. The agent perceives an observation $z$ about the environment (given by a function $Z$) and obtains a reward $r$ (given by a function $R$).}
\label{fig:agente}
\end{figure}

Reinforcement learning (RL) is one important area of machine learning that formalises the interaction of an agent with its environment~\citep{puterman1994markov}. A Markov Decision Process (MDP) can be seen as a model of an agent interacting with the world (see Figure~\ref{fig:agente}), where the agent takes the state $s$ of the world as input and generates an action $a$ as output that affects the world. There is a transition function $T$ that describes how an action affects the environment in a given state. The component $Z$ represents the agent's perception function, which is used to obtain an observation $z$ from the state $s$. In an MDP it is assumed there is no uncertainty in where the agent is. This implies that the agent has full and perfect perception capabilities and knows the true state of the environment (what it perceives is the actual state, $z=s$). The component $R$ is the reward function, the rewards give an indication of the quality of which actions the agent needs to choose. However, the reward function is not always simple to define (for example, it may be stochastic or delayed). Formally,

\begin{definition}[Markov decision process]
An MDP is defined by the tuple $\langle S,A,R,T \rangle$ where $S$ represent the world divided up into a finite set of possible states. $A$ represents a finite set of available actions. The transition function $T : S \times A \rightarrow \Delta(S)$ maps each state-action pair to a probability distribution over the possible successor states, where $\Delta(S)$ denotes the set of all probability distributions over $S$. Thus, for each $s, s' \in S$ and $a \in A$, the function T determines the probability of a transition from state $s$ to state $s'$ after executing action $a$. The reward function $R : S \times A \times S \rightarrow \mathbb{R}$ defines the immediate and possibly stochastic reward that an agent would receive for being in state $s$, executing action $a$ and transitioning to state $s'$. 
\end{definition}

\begin{figure}
\centering
\includegraphics[scale=0.65]{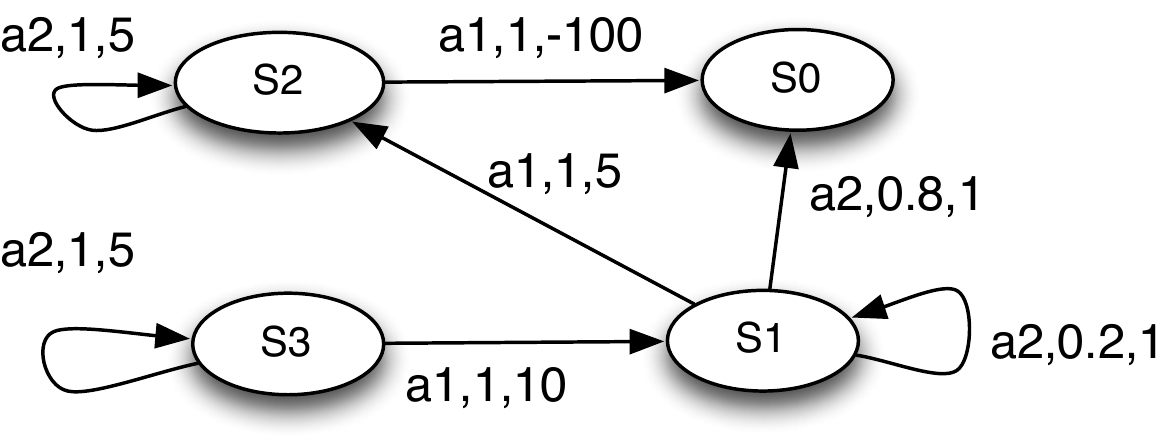}
\caption{ \small A Markov decision process (MDP) with four states $S0, S1, S2, S3$ and two actions $a1, a2$. The arrows denote the tuple: action, transition probability and reward.}
\label{fig:mdp}
\end{figure}

An example of an MDP with 4 states and 2 actions is depicted in Figure~\ref{fig:mdp}, where ovals represent states of the environment. Each arrow has a triplet $a_n,p,r$ representing the action, the transition probability and the reward, respectively. 
   
The key assumption in defining MDPs is that they are \emph{stationary}, i.e., particularly the transition probabilities and the reward distributions do not change in time.\footnote{Formally, an MDP assumes $S, A, R$ and $T$ to be stationary. These sets and function must be unchanged over time, albeit that is compatible with making the action set $A$ stochastic or dependent on the state.} MDPs are adequate models to obtain optimal decisions in environments with a \emph{single} agent. Solving an MDP will yield a policy $\pi: S \rightarrow A$, which is a mapping from states to actions. An optimal policy $\pi^*$ is the one that maximises the expected reward. There are different techniques for solving MDPs assuming a complete description of all its elements. One of the most common techniques is the value iteration algorithm~\citep{Bellman:1957ud} which is based on the Bellman equation:

\begin{equation}
\label{eqn:bellman}
V^{\pi}(s)=  \sum_{a \in A} \pi(s,a)  \sum_{s' \in S}  T(s,a,s') [R(s,a,s') + \gamma V^{\pi}(s')],
\end{equation}

with $\gamma \in [0, 1]$. This equation expresses the \emph{value} of a state which can be used to obtain the optimal policy $\pi^* = \argmax_{\pi} V^{\pi}(s)$, i.e., the one that maximises that value function, and the optimal value function $V^*(s)$.
$$  V^*(s)=  \max_{\pi} V^{\pi}(s) \quad \forall s \in S.$$

%The complexity of solving an MDP depends on the approach, but several methods had been shown to be in $\mathcal{P}$~\citep{Littman:1995ti}.

Finding the optimal policy for an MDP using value iteration requires the MDP to be fully known, including a complete and accurate representation of states, actions, rewards and transitions. However, this may be difficult if not impossible to obtain in many domains. For this reason, RL algorithms have been devised that learn the optimal policy from experience and without having a complete description of the MDP a priori.

An RL agent interacts with the environment in discrete time-steps. At each time, the agent chooses an action from the set of actions available, which is subsequently executed in the environment. The environment moves to a new state and the reward associated with the transition is emitted (see Figure~\ref{fig:agente}). The goal of a RL agent is to maximise the expected reward. In this type of learning the learner is not told which actions to take, but instead must discover which actions yield the best reward by trial and error.

\qlearning~\citep{Watkins:1989uk} is one well known value-based algorithm for RL. It has been devised for stationary, single-agent, fully observable environments with discrete actions. In its general form, a \qlearning agent can be in any state $s \in S$  and can choose an action $a \in A$. It keeps a data structure $\hat Q(s,a)$ that represents the estimate of its expected payoff starting in state $s$, taking action $a$. Each entry $\hat Q(s,a)$ is an estimate of the corresponding optimal $Q^*$ function that maps state-action pairs to the discounted sum of future rewards when starting with the given action and following the optimal policy thereafter. Each time the agent makes a transition from a state $s$ to a state $s'$ via action $a$ receiving payoff $r$, the $Q$ table is updated as follows:

\begin{equation}
\label{eqn:qlearning}
 \hat Q(s,a)= \hat Q(s,a) +\alpha [(r + \gamma \max_{b} \hat Q(s',b))- \hat Q(s,a)]
\end{equation}
with the learning rate $\alpha$ and the discount factor $\gamma \in  [0,1]$ being parameters of the algorithm, with $\alpha$ typically decreasing over the course of many iterations. \qlearning is proved to converge towards $Q^*$ if each state-action pair is visited infinitely often under specific parameters~\citep{Watkins:1989uk}. \qlearning is said to be an \emph{off-policy} method since it estimates the sum of discounted rewards of the optimal policy (aka. target policy) while actually executing an exploration policy (aka. behavior policy) distinct from it.\footnote{The reason that the exploration policy is not the optimal policy is that 1. the optimal policy is not know yet to the agent, and 2. that the action that is most informative is not necessarily the one leading to the highest expected reward.} In contrast, \emph{on-policy} methods refer to algorithms that estimation the value of the executed (exploration) policy. Since the exploration policy is commonly non-stationary, primarily due to the decrease of exploration parameters over time, the target value $Q^*$ to approximate changes with it, making it more intricate to provide convergence results.\footnote{Convergence proofs for on-policy methods usually require more details to be specified than for off-policy algorithms~\citep{Singh:2000ur,VanSeijen:2009fe}.} One classic on-policy algorithm is SARSA (state$_{t}$, action$_{t}$, reward$_{t}$, state$_{t+1}$, action$_{t+1}$)~\citep{suttonBarto1998} which uses a variation of Equation~(\ref{eqn:qlearning}):
\begin{equation}
\label{eqn:sarsa}
 \hat Q(s,a)= \hat Q(s,a) +\alpha [(r + \gamma \hat Q(s',a'))- \hat Q(s,a)].
\end{equation}
%
%\emph{expected SARSA}, which computes an expectation over all actions available in $s'$:
%\begin{equation}
%\hat Q(s,a)= \hat Q(s,a) +\alpha [(r + \gamma  \sum_a \pi(s',a)\hat Q(s',a))- \hat Q(s,a)]
%\end{equation}
%provides convergence guarantees to the optimal policy \citep{VanSeijen:2009fe}. Moreover, expected SARSA can be viewed as an on-policy version of Q-learning.

By using \qlearning it is possible to learn an optimal policy without knowing $T$ or $R$ beforehand, and even without learning these functions~\citep{Littman:1996ts}. For this reason, this type of learning is known as \emph{model free} RL. In contrast, \emph{model-based} RL aims to learn a model of its environment, specifically approximating $T$ and $R$. Such models are then used by the agent to predict the consequences of actions before they are taken, facilitating planning ahead of time. One example of this type of algorithms is Dyna-Q~\citep{suttonBarto1998}.

\paragraph{Exploration vs exploitation}
\label{sec:sampleComplexity}

Similar to multi-armed bandits, in RL one main concern is to develop algorithms that balance exploration and exploitation well. However, in contrast to bandits where algorithms are evaluated in terms of regret, the RL community has proposed different measures to determine \emph{efficient exploration}. An important concept is the \emph{sample complexity} \citep{Vapnik:1998uq}, which was first defined in the context of supervised learning. Loosely speaking, sample complexity is the number of examples needed to bring the estimate of a target function within a given error range. \cite{Kakade:2003vv} studied sample complexity in a RL context.  Consider an agent interacting in an environment. The steps of the agent can be roughly classified into two categories: steps in which the agent acts near-optimally as ``exploitation'' and steps in which the agent is not acting near optimally as ``exploration''. Subsequently, it is possible to see the number of times in which the agent is not acting near-optimally as the \emph{sample complexity of exploration}~\citep{Kakade:2003vv} for which some algorithms have guarantees~\citep{Brafman:2003vk}.

%Classical RL considers one single agent in a stationary environment. %Before discussing the various extensions to multiagent RL that draw on the insights from MAB, RL and game theory, \reviseq{check previous sentence. where do you do this in the paper? This sounds like you will have a section on this topic, while I assume you refer to there being some multi-agent rl algorithms in section 5.} 
%\revise{
Before formalizing the problem of learning in multiagent environments (Section~\ref{sec:learningMAS})
%}
we will discuss game theory in the next section, as it is a classical area that addresses the interaction, reasoning and decision-making of multiple agents in strategic conflicts of interest.

%\reviseq{- I like that the authors ended section 2.1 with a paragraph discussing what this approach is appropriate for, but that it doesn’t necessarily deal effectively with non-stationarity. I think they could do the same for sections 2.2 and 2.3 (perhaps I just missed it). While the rest of the paper addresses the topic, I think it makes sense here to highlight the problem at the end of these subsections.}

%\revise{To summarise, reinforcement learning is appropriate for learning an acting policy in fully observable environments where the state transitions are Markovian (e.g., single-agent environments)...For more details we refer the reader to recent books and surveys on the topic~\citep{sutton2018reinforcement,franccois2018introduction}.}

\subsection{Game theory}
\label{sec:prelim:gametheory}

Game theory studies decision problems when several agents interact~\citep{Fudenberg:1991vw}. The terminology in this area is different, agents are usually called \emph{players}, a single interaction between players is represented as a \emph{game}, and rewards obtained by the players are called \emph{payoffs}.

The most common way of presenting a game is by using a matrix that denote the utilities obtained by each agent, this is the normal-form game. 

\begin{definition}[Normal-form game] \label{def:normal-form}
$A$ (finite, $\mathcal{I}$-person) normal-form game $\Gamma$, is a tuple $ \langle \mathcal{N},A,u \rangle$, where:
\item $\mathcal{N}$ is a finite set of $\mathcal{I}$ players, indexed by $i$;
\item $A = A_1 \times \dots \times A_\mathcal{I}$, where $A_i$ is a finite set of actions available to player $i$. Each vector $a=(a_1, \dots, a_\mathcal{I}) \in A$ is called an action profile;
\item$ u= (u_1, \dots, u_\mathcal{I})$ where $u_i: A  \mapsto \mathbb R$ is a real-valued utility or payoff function for player $i$. 
\end{definition}

\begin{table}
\caption{ \small The normal-form representation of the prisoner's dilemma game. Each cell represents the utilities given to the players (left value for $\mathcal{A}$ and right one for $\mathcal{O}$), $r_{pd},t_{pd},s_{pd},p_{pd} \in \mathbb{R}$ where the following conditions must hold $t_{pd} > r_{pd} > p_{pd} > s_{pd}$  and $2r_{pd} > p_{pd}+s_{pd}$. }
\begin{center}
\begin{tabular}{cccc} \toprule
\multicolumn{2}{c}{}	        	 	 &	\multicolumn{2}{c}{Player $\mathcal{O}$}\\ 
\multicolumn{2}{c}{}			  & \emph{cooperate} & \emph{defect} \\ \hline
		\multirow{2}{*}{Player $\mathcal{A}$}    & \emph{cooperate} & $r_{pd},r_{pd}$ & $s_{pd},t_{pd}$ \\ 
							& \emph{defect}	 & $t_{pd},s_{pd}$ & $p_{pd},p_{pd}$ \\ 
\bottomrule
\end{tabular}
\end{center}
\label{tab:prisionersdilemma}
\end{table}

For example, Table~\ref{tab:prisionersdilemma} shows a two-action two-player game, known as the \emph{Prisoner's Dilemma} (PD). Each row corresponds to a possible action for player $\mathcal{A}$ and each column corresponds to a possible action for player $\mathcal{O}$. Player's payoffs are provided in the corresponding cells of the joint action, with player $\mathcal{A}$'s utility listed first. In the example, each player has two actions \{\emph{cooperate, defect}\}. A \emph{strategy} specifies a method for choosing an action. One kind of strategy is to select a single action and play it, this is a \emph{pure} strategy. In general, a \emph{mixed} strategy specifies a probability distribution over actions. 
\begin{definition}[Mixed strategy]
Let $(\mathcal{I},A,u)$  be a normal-form game, and for any set $X$, let $\Delta(X)$ be the set of all probability distributions over $X$, then the set of mixed strategies for player $i$ is $\mathcal{S}_i=\Delta(A_i)$ 
\end{definition}
In this context, it is important to define what is a good strategy, i.e., the best response. 
\begin{definition}[Best response]
Player $i$'s best response to the strategy profile $s_{-i}$ is a mixed strategy $s_i^* \in \mathcal{S}_i$ such that $u_i(s_i^*,s_{-i}) \ge u_i(s_i,s_{-i}) $ for all strategies $s_i \in \mathcal{S}_i$.
\end{definition}
\noindent where $s_{-i}= s_1, \dots, s_{i-1},s_{i+1}, \dots,s_n $ represents the strategies of all players except $i$. Thus, a \emph{best response} for an agent is the strategy (or strategies) that produce the most favourable outcome for a player, taking other players' strategies as given.  Another common strategy is the \emph{minimax strategy} that ensures a \emph{security level} for the player. 

\begin{definition}[Minimax Strategy.]
Strategy that maximizes its payoff assuming the opponent will make this value as small as possible.
\end{definition}

\begin{definition}[Security level]
The \emph{security level} is the expected payoff a player can guarantee itself using a minimax strategy.
\end{definition}

In single-agent decision theory, the notion of \emph{optimal strategy} refers to the one that maximises the agent's expected payoff for a given environment. In multiagent settings the situation is more complex, and the optimal strategy for a given agent may now vary, since the best response strategy depends on the choices of others. In order to draw conclusions on the joint behavior in games, game theory has identified certain subsets of outcomes, called solution concepts, such as the \emph{Nash equilibrium} (NE). Suppose that all players have a fixed strategy profile in a given game, if no player can increase its utility by \emph{unilaterally} changing its strategy, then the strategies are in Nash equilibrium. Formally it is defined by:

\begin{definition}[Nash equilibrium;~\citealp{Nash:1950vo}]
A set of strategies $s=(s_1,\dots ,s_n)$ is a Nash equilibrium if, for all agents $i$, $s_i$ is a best response to $s_{-i}$.
\end{definition}

%\reviseq{- In my opinion, the discussion of Nash equilibrium in section 2.3 is inadequate and misleading. For example, an appeal to “the” NE of the PD when you are talking about multi-agent learning (which entails that there is repeated interaction of some form) means that the game has more than one NEs (of the repeated game). Depending on the nature of the repeated interaction and the game structure, there could be infinite NE (folk theorem). While “one-shot” NEs are NE of the repeated game, talking about one such NE as “the” NE of the PD is misleading. Sorry, I’ll get off my soap box now. :-) This isn’t a main issue, but my opinion is that the current treatment in the paper is misleading.}

Even when it is proved that in every game exists a Nash equilibrium, this solution concept has limitations. One problem is that there may be multiple equilibria in a game, and it is not an easy task to select one~\citep{harsanyi1998general}. Also several experiments involving humans have shown that that people usually do not follow the actions prescribed by the theory~\citep{Kahneman:1979wl,Risse:2000wb,Goeree:2001vy,Camerer:2003vb}. %\revise{We could remove this part "Another complaint is that, there exist many games in which the Nash equilibrium does not guarantee the maximum utility for all players. For example, in the PD game (see Table~\ref{tab:prisionersdilemma}) the Nash equilibrium occurs when both players play defect, however, many experiments have shown that people cooperate in similar situations~\citep{Camerer:2003vb}."}

\paragraph{Extensive-form games}

Another common representation for games is the \emph{extensive-form} in which it is easier to describe the sequential structure of the decisions (for example, this is useful to represent poker games). Commonly, the game is described as a tree where nodes represent actions taken by the players. Extensive-form games can be finite or infinite-horizon (regarding the length of the longest possible path), with observable or non-observable actions and with complete or incomplete information (observability of the opponent payoffs)~\citep{Fudenberg:1991vw}. Most of the games represented in extensive form can be converted into a normal-form representation, however, this generally results in a matrix which is exponential in the size of the original game. For this reason, it is common to find a solution in the original game tree.

\paragraph{Repeated and stochastic games}
\label{sec:prelim:rg}

Previous concepts (e.g., best response, Nash equilibrium) were defined for one-shot games (one single interaction), however, it could be the case that more than one decision has to be made. For example, repeating the same game, or having a set of possible games. 

\begin{definition}[Stochastic game]
A stochastic game (also known as a Markov game) is a tuple $(S,\mathcal{N},A,T,R)$, where:
$S$ is a finite set of states, $\mathcal{N}$ is a finite set of $\mathcal{I}$ players, $A= A_1 \times \dots \times A_\mathcal{I}$ where $A_i$ is finite set of actions available to player $i$, $T:S \times A \times S \rightarrow \mathbb{R}$ is the transition probability function; $T(s,a,\hat{s})$ is the probability of transitioning from state $s$ to state $\hat{s}$ after action profile $a$, and $R= r_1, \dots, r_\mathcal{I}$ where $r_i: S \times A \rightarrow \mathbb{R}$ is a real valued payoff function for player $i$.  
\end{definition}
In a stochastic game, agents repeatedly play games (states) from a collection. The particular game played at any given iteration depends probabilistically on the previous played game (state), and on the actions taken by all agents in that game~\citep{Shoham:2008vg}. 
\begin{definition}[Repeated game]
A repeated game is a stochastic game in which there is only one game (called stage game).
 \end{definition}
 
\begin{figure}
\begin{center}
\begin{tabular}{cc}
\subfigure[]{
\includegraphics[scale=0.60]{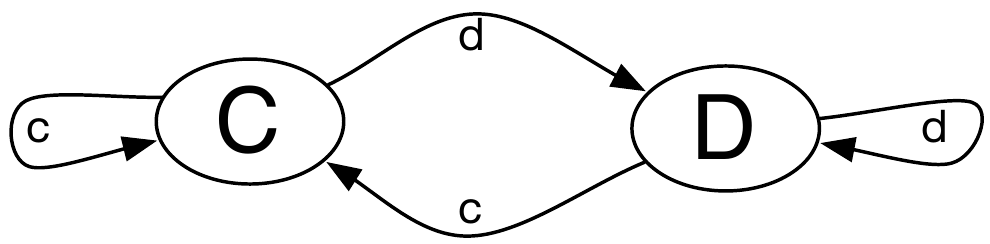}
}
&
\subfigure[]{
\includegraphics[scale=0.60]{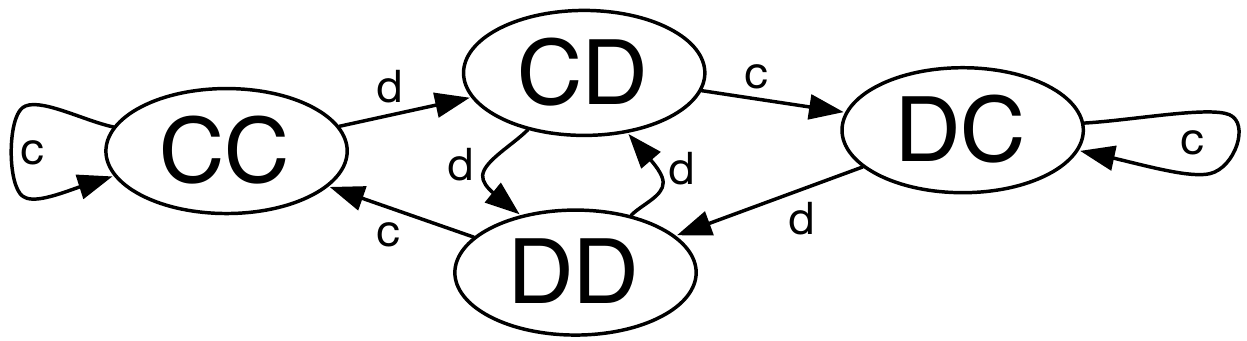}
}
\end{tabular}
\caption[The automata that describe TFT and Pavlov strategies.]{ \small (a) The automata that describes the TFT strategy, depending of the opponent action (c or d) it transitions between the two states C and D. (b) The automata describing Pavlov strategy, it consists of four states formed by the last action of both agents (CC, CD, DC, DD).}
\label{fig:automatas}
\end{center}
\end{figure}

To exemplify a repeated game, recall the prisoner's dilemma presented in Table~\ref{tab:prisionersdilemma}. Repeating the game for a number of rounds results in the iterated prisoner's dilemma (iPD), which has been the subject of different experiments and for which there are diverse well-known strategies. A successful strategy which won Axelrod's tournament\footnote{Robert Axelrod held a tournament of various strategies for the iterated prisoner's dilemma. Strategies were run by computers. In the tournament, programs played games against each other and themselves repeatedly.} is called Tit-for-Tat (TFT)~\citep{Axelrod:1981vw}; it starts by cooperating, and does whatever the opponent did in the previous round: it will cooperate if the opponent cooperated, and will defect if the opponent defected. Another important strategy is called Pavlov, which cooperates if both players performed the same action and defect whenever they used different actions in the past round. The finite-state machines describing TFT and Pavlov are depicted in Figure~\ref{fig:automatas}. It should be noticed that these strategies can be described in terms of the past actions and therefore do not depend on the time index; they are stationary strategies. 

%\reviseq{- I like that the authors ended section 2.1 with a paragraph discussing what this approach is appropriate for, but that it doesn’t necessarily deal effectively with non-stationarity. I think they could do the same for sections 2.2 and 2.3 (perhaps I just missed it). While the rest of the paper addresses the topic, I think it makes sense here to highlight the problem at the end of these subsections.}

%\revise{Game theory is appropriate for...}

\textcolor{black}{
Having presented the formal models of multi-armed bandits, reinforcement learning and game theory, the next Section highlights the challenge of non-stationarity in multiagent systems, followed by a new framework for this setting. Moreover, we present our proposed categorisation on how algorithms cope with non-stationary behaviour.}

% !TEX root = survey.tex

\section{Learning in multiagent environments}
\label{sec:learningMAS}

\textcolor{black}{
The following subsection pinpoints where and how the main challenge of non-stationarity arises in multiagent environments. This provides a crisp basis problem definition against which the approaches of algorithms can be positioned. Next, we present a new abstract framework for multiagent learning algorithms that naturally models (and emphasizes) three components of accounting for other reasoning agents. Finally, we present a taxonomy of multiagent learning algorithms, aligned with assumptions they make in light of this framework.
}

\subsection{The problem}

Learning in a multiagent environments is inherently more complex than in the single-agent case, as agents interact at the same time with environment and potentially with each other~\citep{Busoniu:2008bo}. Transferring single-agent algorithms to the multiagent setting is a natural heuristic approach (see Section~\ref{sec:categories}) -- even if assumptions under which these algorithms were derived are violated. In particular the \emph{Markov property}, denoting a stationary environment, does not hold, thus invalidating guarantees derived for the single-agent case~\citep{Tuyls:2012up}. Since this approach of applying single-agent algorithms ignores the multiagent nature of the setting entirely, it can fail when an opponent may adapt its choice of actions based on the past history of the game~\citep{Shoham:2007vw}.

In order to expose why multiagent domains are non-stationary from agents' local perspectives, consider a stochastic game $(S,\mathcal{N},A,T,R)$. Given a learning agent $i$ and using the common shorthand notation $\bm{-i} = \mathcal{N} \setminus \{ i \}$ for the set of opponents, the value function now depends on the joint action $\bm{a} = (a_i, \bm{a_{-i}})$, and the joint policy $\bm{\pi}(s, \bm{a}) = \prod_j \pi_j (s, a_j) $: 

\begin{equation}
\label{eqn:bellmanMAS}
V^{\bm{\pi}}_{i}(s)=  \sum_{\bm{a} \in A} \bm{\pi}(s,\bm{a})  \sum_{s' \in S}  T(s,a_i,\bm{a_{-i}},s') [R(s,a_i,\bm{a_{-i}},s') + \gamma V_{i}(s')].
\end{equation}
Consequently, the optimal policy is a best response dependent on the other agents' policies,
\begin{align*}
&\pi_i^*(s,a_i,\bm{\pi_{-i}}) = BR_i(\bm{\pi_{-i}}) = \argmax_{\pi_i} V^{(\pi_i, \bm{\pi_{-i}})}_{i}(s)\\
&= \argmax_{\pi_i} \sum_{\bm{a} \in A} \pi_i(s, a_i) \bm{\pi_{-i}}(s,\bm{a_{-i}})  \sum_{s' \in S}  T(s,a_i,\bm{a_{-i}},s') [R(s,a_i,\bm{a_{-i}},s') + \gamma V^{(\pi_i, \bm{\pi_{-i}})}_{i}(s')].
\end{align*}
Specifically the opponents' joint policy $\bm{\pi_{-i}}(s,\bm{a_{-i}})$ %\revise{
could be non-stationary, (for example when opponents' are learning) thus becoming the parameter of the best response function. If the opponents' are not learning, e.g., they are using a stochastic policy, then the environment is Markovian and single-agent learning algorithms suffice.%}
% TODO: We might want to add pointers to 'if pi_-i was stationary, this could be solved with value or policy iteration'.

Next, we propose a general framework for multiagent learning algorithms, separating three steps of modelling opponents' behaviour to tackle the problem of non-stationary opponent policies.

\subsection{A new framework for multiagent learning algorithms}
\label{sec:newFramework}

Before going into the formal definitions of the abstract concepts, consider an intuitive description of the three components of our proposed framework:
\begin{itemize}
\item \emph{Policy generating functions} $\tau \in \mathcal{T}$, describe how an opponent $j$ obtains its policy $\pi_j$.
\item \emph{Belief} $\beta_j$, i.e., a probability distribution over $\tau$, measures an agent's belief about each opponent's  reasoning.
\item \emph{Influence function} $\theta$ partitions beliefs according to equivalent best responses.
\end{itemize}

Some definitions are required to describe each component in detail: let $h_t=(z_0,z_1, \dots,z_t)$ denote the \emph{observation history} of $t$ observations, and let $H^t$ be the set of all possible histories of this length. Note that observations in the stochastic game are given by \emph{state, action} sequences, but this more general representation also subsumes models of partial observability, such as POMDPs. While rewards are commonly treated separately in the literature, they may simply be added as part of the observation history in our model. Some work may presume or learn a model of reward functions, as in POMDPs for the agent~\citep{KaelblingLP:1998vs} or the \emph{frame} of I-POMDPs for opponents~\citep{Gmytrasiewicz:2005un}, which is equally compatible with our model.

%let $h_t=(s_0,s_1,s_2 \dots,s_t)$ denote the \emph{state history} of $t$ states, and let $H_s^t$ be the set of all possible histories of this length. Similarly, let $h_t=(a_0,a_2,a_2 \dots, s_t)$ denote the \emph{action history} of $t$ states, and let $H_a^t$ be the set of all possible histories of this length.

\begin{definition}[Policy generating functions]
A policy generation function (PGF) $\tau$ maps the history of observations into a policy $\pi$, 
$$\tau: H^t \rightarrow \Pi$$
\end{definition}
On the one hand, this definition could be extended to the stochastic case $\tau: H^t \rightarrow \Delta(\Pi)$, where $\Delta(\cdot)$ indicates the simplex function, i.e., here denoting a probability distribution or probability mass function over policies. However, this complexity appears unnecessary for the exposition we aim for in this section. On the other hand, the composition of $\tau$ and $\pi$ could be chosen as an alternative definition, % TODO: and has been in related work [cite someone]
thus mapping histories directly to actions. 

In contrast to alternative definitions, deterministic policy generating functions are a particularly relevant category since they capture memory-bounded models with hidden states, while maintaining the structure of policies. This enables additional assumptions over the rate of change in policies, or the set of policies that are (re-) visited by the algorithm. Such models subsume learning algorithms, e.g., \qlearning~\citep{Watkins:1989uk}, weight matrices describing neural networks~\citep{Bengio:2009kb}, and MDPs, e.g., mapping a sliding window of the histories to an action or policy, as finite state automata over a predefined set of policies~\citep{Banerjee:2005wq,Chakraborty:2013ii}.

The PGFs capture the adaptation dynamics of agents, and research articles derive insights related to learning algorithms within a scope delimited by an implicitly or explicitly defined set of PGFs for any opponents. One of the more general assumptions is given by the \emph{frame} defined in I-POMDPs~\citep{Gmytrasiewicz:2005un}, which assumes further structure on the PGFs, such as ascribing rewards and optimality criteria to opponents. Our taxonomy below employs PGF assumptions as a main criterion for classifying algorithms.

The next step is to define how the agent uses those PGFs. Note that observations are local to each agent, i.e., an agent $i$ can only infer another agent's local perceived observation history $h_j$ by a probability distribution $p(h_j | h_i)$ using its own observations $h_i$ together with any available a priori knowledge, e.g., about the structure of the game. In stochastic games, \emph{state, action} sequences are joint observations,\footnote{Stochastic games usually assume that agents have complete information about the state of the game, a more general model are partially observable stochastic games (POSGs)~\citep{Bernstein:2004uu}.} thus $h_j = h_i$ if rewards are treated separately.

\begin{definition}[Belief]
A belief $\beta \in \mathcal{B}$ indicates for each opponent $j$ the likelihood $\beta_j(\tau|h_j)$ for each policy generating function $\tau$ given opponent experience $h_j$.
% \[ 
  %\beta: \mathcal{N} \rightarrow \Delta(\tau).
%   \beta_j(\tau|h_j) \quad h_j \sim p(h_j|h_i).
% \]
\end{definition}
%\comment{We may want to insert belief computation here, and specifically notation $\beta_j(\tau | h_t)$, using the assumed connection between $h_j$ and $h_i$ above.}
% 
Since $h_j$ is local information, it must be inferred from agent $i$'s observations, i.e., $h_j \sim p(h_j|h_i)$.
If the presumed set $\mathcal{T}$ gives rise to distinguishable policies, then the belief may identify each opponent's PGF in a crisp belief (assigning probability one to a specific $\tau$ for each opponent). Even if unique identification is not possible, this poses a classification task, and a unique assignment may be used as an approximation of the belief. On the other hand, full belief representations over multiple $\tau$ for each opponent are common in Bayesian reasoning~\citep{Ghavamzadeh:2015jg}.

The last step defines how the belief could be filtered (e.g., to reduce complexity) by means of an influence function.

\begin{definition}[Influence function for multiagent learning]
The co-domain of the influence function $\theta$ over the belief is a k-dimensional influence space $\Theta$:
$$ \theta: \mathcal{B} \rightarrow \Theta.$$
%\comment{Influence space should be a more general space, possibly continuous and discrete.}

\end{definition}
Assumptions about the influence function may significantly alter the complexity of the algorithm, and the validity of such assumptions differentiates whether resulting model insights hold or reduce to heuristic approaches; below we provide some examples.
\begin{itemize}
\item In single-agent learning, the assumption is that $\theta$ maps onto a singleton set. 
\item On the opposite side of the spectrum, taking the identity function as $\theta$ is equivalent to not modelling $\theta$ at all, thus also not limiting the validity at this step. 
\item However, imposing---or learning---a structure of $\theta$ would cluster equivalent best responses~\citep{Bard:2015ta} and may lead to more sample efficient learning of best response approximations. One example instantiation of the influence function may encode abductive reasoning by mapping mixed beliefs to crisp classifications, as mentioned in the above discussion of beliefs. 
\item Furthermore, in symmetric games with distinguishable policies, $\theta$ may encode the strategy histogram (counting players for each $\tau$), as by definition the payoffs of a player only depend on the strategies employed, and not on who is playing them. This structure is used in heuristic payoff tables to compress utility representations and corresponding best response mappings~\citep{Walsh:2002ur}.
\end{itemize}
Overall, an influence function typically reduces the complexity, either as a lossless compression or as a heuristic to reduce the set of best responses and the computational complexity of deriving them.

\begin{definition}[Best response in multiagent learning] A multiagent learning algorithm computes the best response to the influence state of its belief, given an a priori assumed $\mathcal{T}$:
%$$BR_i(\theta) = \pi^*_i(s,a,\theta) = BR_i(\bm{\pi_{-i}} | \bm{\pi_{-i}} \sim \beta(\tau(H_s^t, H_a^t))),$$
$$BR_i(\hat \theta) = \pi^*_i(s,a,\hat \theta) =
BR_i\left(\bm{\pi_{-i}} | \pi_{j} \sim \beta_j(\tau|h_j),h_j \sim p(h_j|h_i) \right),$$
\end{definition}
for any $\beta$ that satisfies $\theta(\beta) = \hat \theta$. % slight abuse of notation as function and specific parameter

This new framework maintains agent independence by mutual non-observability of individual policies, and inherently models agent autonomy by the independent choice of best response policies.

\subsection{Taxonomic Categories: Environment, Opponent and Agent}
\label{sec:categories}

%Before distilling each algorithm in detail, in this section 
Now, we present a taxonomy in terms of the environment (observability) and the opponent characteristics (learning capabilities). Then, we provide an overview of the proposed categories of how algorithms deal with non-stationarity. 

\subsubsection{Environment: observability}

%\reviseq{Section 3.1
%The authors list four different categories of “observability.” I find these categories to be too shallow (incomplete). What about the ability to observe the states of the world (rather than just the actions)? What about the ability to observe the state of the “opponent?” For example, one might be able to observe whether their opponents are happy/satisfied or not. One might be able to observe what algorithms opponents are using. Additionally, what about communication? How much can I communicate with my opponent? Many of the algorithms designed for multi-agent learning ignore explicit communication. This limiting assumption is worth thinking about, and work is now emerging that challenges this assumptions. However, communication can help us deal with some forms of non-stationarity.
%In general, these questions are very vast. Thus, I can understand a desire to remain simplistic. However, for a paper that is about assumptions (and helping us to think about the assumptions we make), the discussion in this section is much too simplistic.}
One crucial aspect that provides information on how to tackle a learning problem is observability of actions and rewards, for both the learning agent and the opponent.  
Depending on the restriction of the domain, there are four categories in increasing order of observability.
\begin{description}
\item [Local reward.] The most basic information that an algorithm commonly observes are its own immediate rewards. 
\item [Opponent actions.] Most algorithms also assume that is possible to observe the opponent actions (but not the their rewards).
\item [Opponent actions and payoffs.] Some algorithms assume to observe the action and also the actual payoffs of the opponents (which may hold more naturally in cooperative scenarios).
\item [Complete a priori knowledge.] Similar to the previous category algorithms observe rewards and actions, however, in this category the algorithms know from start the complete reward function.
\end{description}

\subsubsection{Opponent: adaptation capabilities}
\label{sec:opponentbehavior}

%\textcolor{red}{
%Opponent policy set.?
%}
%\revise{The categories below could be named 'Adaptation temporal types':}

The capability of the opponent to adapt and change its behaviour provides another source of important information to be used while learning. Roughly, we distinguish three categories:
\begin{description}
\item [No adaptation. $\forall h^t: \tau(h^t) = \pi$] These are opponents that follow a stationary strategy  during the complete period of interaction.
\item [Slow adaptation. $\exists \epsilon << 1, \forall t: d\left(\tau(h^{t+1}), \tau(h^t) \right) < \epsilon$] These opponents show non-stationary behaviour. However, it is a limited adaptation, for example providing bounds to the possible change in the current strategy between rounds. Candidate metrics are Manhattan distance $d_1$ or the average Jensen-Shannon distance over all states, which with base 2 logarithm is bound to $[0, 1]$: $d\left(\tau(h^{t+1}), \tau(h^t) \right) \hat = \frac1{|S|} \sum_s \sqrt[]{JSD\left(\tau_s(h^{t+1}) || \tau_s(h^t) \right)}$.
\item [Drastic or abrupt adaptation. ] If the above assumptions are not in place, non-stationary opponents may show abrupt changes in their behaviour, for example changing to a different strategy (no limits) from one step to the next. 
\end{description}

%\revise{We could consider another categorisation for
%'Adaptation causes':
%-learning
%-randomness
%-?
%}

%\reviseq{Section 3.2
%The authors list three categories of the capability of opponents to adapt. These categories seem to focus solely on how fast one’s opponents might adapt (if they adapt at all). Again, I find these categories to not adequately categorize non-stationarity. In addition to knowing how fast opponent(s) might adapt, one *must* (or at least researchers do) also make assumptions about what causes the opponents’ behaviors to change. Where does the non-stationarity come from? There’s an important difference between non- stationarity driven by an opponent adapting to the behavior of others (in order to achieve their own goals) as opposed to non-stationarity governed by other changes in opponents (examples, a robot is getting old and its wheel motors are starting to slip more often; or the power goes out; etc.). Also, will the non- stationarity change over time?
%Granted, quantifying relevant attributes related to non-stationarity may be quite complex (which is probably why people have been working on it for a long time, but have not adequately categorized it). However, since the paper is focused on the problem of non-stationarity, it is my opinion that more than a surface-level treatment is warranted.}

\subsubsection{Agent: dealing with non-stationarity}

Previous surveys have proposed different ways to categorise algorithms in multiagent systems such as: team vs concurrent learning~\citep{Panait:2005wj}; temporal difference, game theory and direct policy search~\citep{Busoniu:2010ft}; model-based, model-free and regret minimization~\citep{Shoham:2007vw}; and  joint action, gradient, Nash and other learners (\citealp[see][chap. 10]{Weiss:2013ue}). There are also some previous categories for the type of learning used: multiplied, divided and interactive~\citep{Tuyls:2012up}; and independent, joint-action and gradient ascent~\citep{Bloembergen:2015ei}. 

Another group of works have proposed properties that MAS algorithms should have:~\cite{Bowling:2002vva} propose rationality and convergence. The former needs the learning algorithm to converge to a stationary policy that is a best-response to the other players’ policies if the other players’ policies converge to stationary policies; the latter refers to the need of the agent to necessarily converge to a stationary policy. \cite{Powers:2004wd} proposed: targeted optimality, compatibility and safety. The first one needs the agent to achieve within $\epsilon$ of the expected value of the best response to the actual opponent. Compatibility needs the algorithm to achieve at least within $\epsilon$ of the payoff of some Nash equilibrium that is not Pareto dominated by another NE (during self-play), and safety needs the agent to receive at least within $\epsilon$ of the security value for the game. \cite{Crandall:2011dt} proposed: security, coordination and cooperation. Security refers to long-term average payoffs meet a minimum threshold, coordination refers to the ability to coordinate behaviour when associates share common interests, and cooperation is the ability to make compromises that approach or exceed the value of the Nash bargaining solution~\citep{Nash:1950hk} in games of conflicting interest.

In contrast with previous works, we propose another view focused on how algorithms deal with non-stationary behaviour. We propose five categories in increasing order of sophistication which we summarize as follows:
\begin{description}
\item [1. Ignore.] The most basic approach which assumes a stationary environment.
\item [2. Forget.] These algorithms adapt to the changing environment by forgetting information and at the same time updating with recent observations, usually they are model-free approaches. 
\item [3. Respond to target opponents.] Algorithms in this group have a clear and defined  target opponent in mind and optimize against that opponent strategy.
\item [4. Learn opponent models.] These are model-based approaches that learn how the opponent is behaving and use that model to derive an acting policy. When the opponent changes they need to update its model and policy.
\item [5. Theory of mind.] These algorithms model the opponent assuming the opponent is modelling them, creating a recursive reasoning.
\end{description}

%\reviseq{First, I can think of at least two other methods that have been used to deal with non-stationarity off the top of my head that do not appear to be captured by these five categories. I suspect there are others. One of these techniques is epitomized by the work of Littman and Stone on “leader algorithms.” Other work on Stackelberg equilibrium hints at the same technique. One can effectively deal with some forms of non- stationarity in an opponent by computing and always playing stationary (nonchanging) strategy that will cause an adaptive opponent (and one must quantify how the opponent adapts) to adapt to “me” rather than “I” having to adapt to them.}
%\revise{Where do we put leader algorithms? I would say target opponent. They assume the opponent is a learning algorithm which is going to adapt to them. }

%\reviseq{The second technique is epitomized by work in aspiration learning (Stimpson and Goodrich, Karandikar et al., Herb Simon) which rather than learning an “opponent model” learns instead an aspiration level that essentially quantifies a reasonable payoff that I should be able to achieve (and then finds a strategy that achieves that level). This paradigm shift in *what to learn* appears to be a very valid mechanism for dealing with some forms of non-stationarity}
%\revise{This aspiration level algorithm could be positioned in the ingore because it kind of not learn a model of the opponent, however I think we should considered into the forget category because it does adapt in a way}%

Note that this order is according to the sophistication in terms of complexity in assumptions and approach---it is throughout possible that the elegance of solutions does not follow this ordering. Moreover, we acknowledge that some algorithms could fit in more than one category. To better understand the high level behaviour of these categories, the next section presents an illustrative example using a simple domain.

\section{Illustrative Example - Iterated Prisoner's Dilemma}
\label{sec:example}

\begin{figure*}
\begin{center}
\subfigure[Ignore: $\mathcal{A}$ assumes an opponent which is stationary ($S$) for the complete interaction period. Examples are \qlearning~\citep{Watkins:1992ve} and fictitious play~\citep{Brown:1951vc}.]{
\includegraphics[scale=0.40]{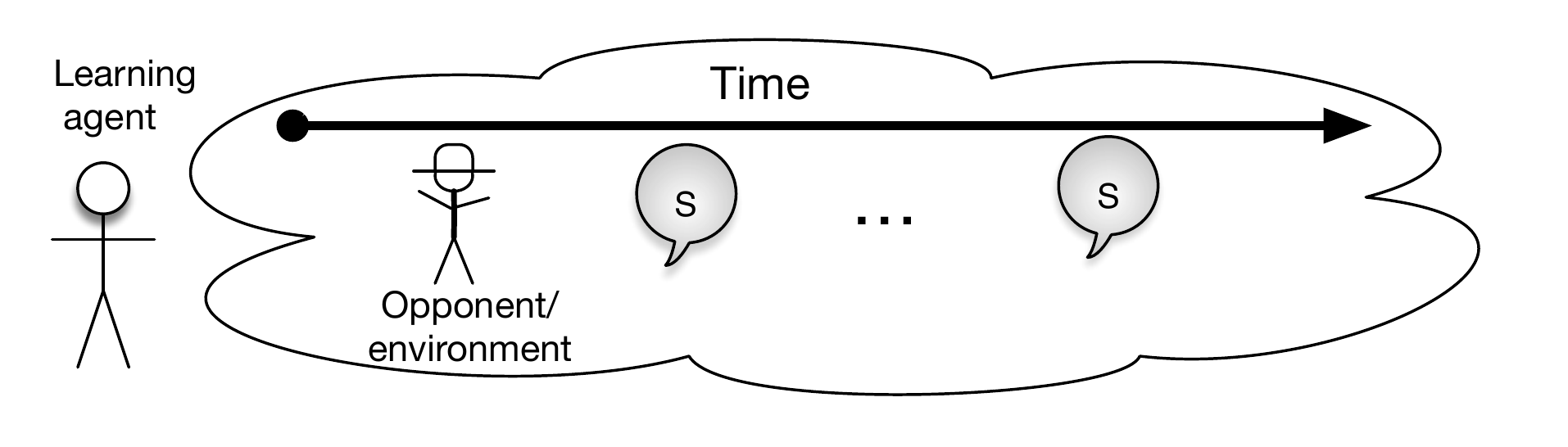}
\label{fig:categories:ignore}
}
\vspace{-0.1cm}
\subfigure[Forget: $\mathcal{A}$ learns an initial strategy ($S1$) which is continually updated ($S1', S1'',\dots$) with recent observations, one example is WoLF-PHC~\citep{Bowling:2002vva}.]{
\includegraphics[scale=0.40]{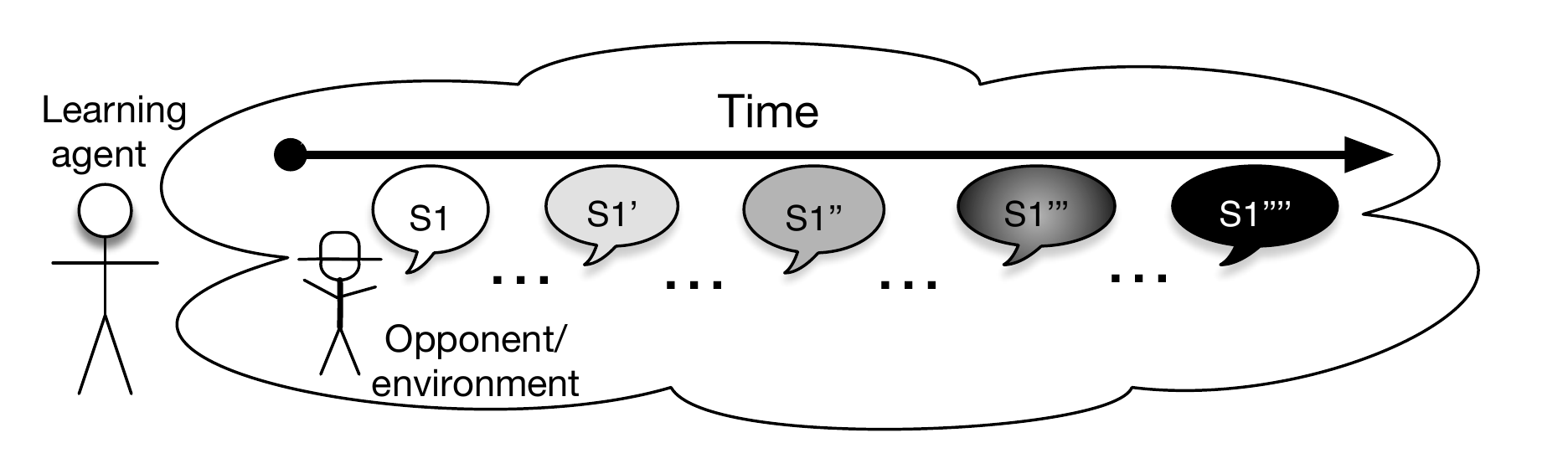}
\label{fig:categories:forget}
}
\vspace{-0.1cm}
\subfigure[Respond to target opponents: One example is Minimax-Q~\citep{Littman:1994ta} where the learning agent assumes the opponent tries to minimise the rewards.]{
\includegraphics[scale=0.40]{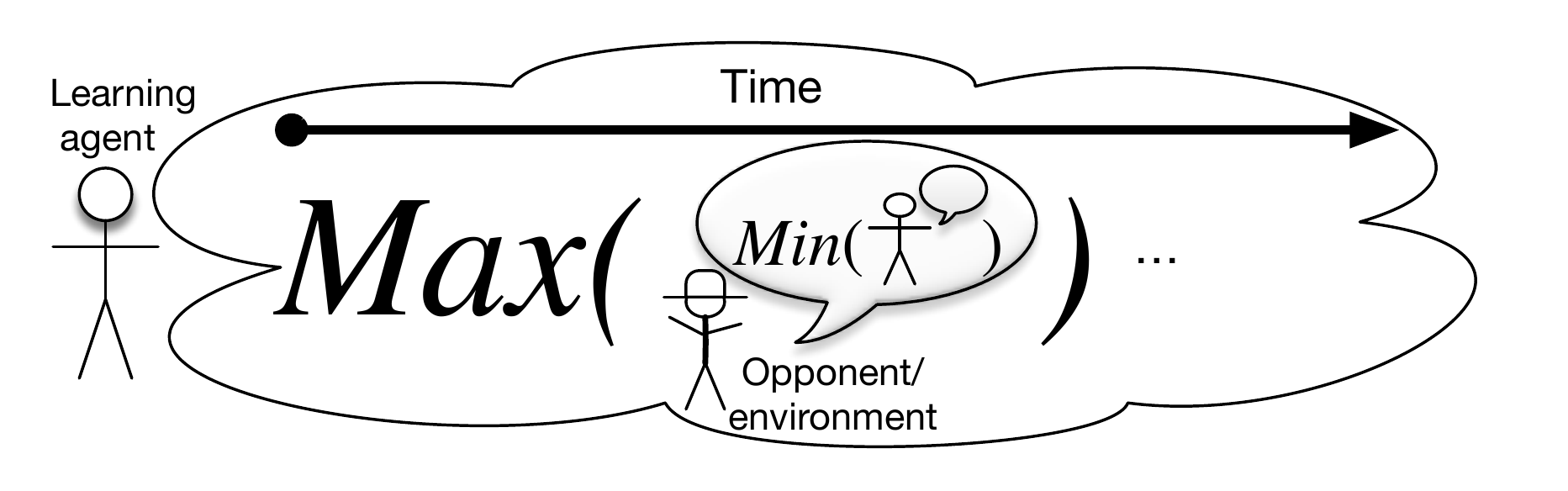}
\label{fig:categories:target}
}
\vspace{-0.1cm}
\subfigure[Learn: $\mathcal{A}$ learns a model of the opponent strategy $(S?)$ and derives an acting policy; opponent changes are infrequent, e.g., RL-CD~\citep{DaSilva:2006uw}.]{
\includegraphics[scale=0.40]{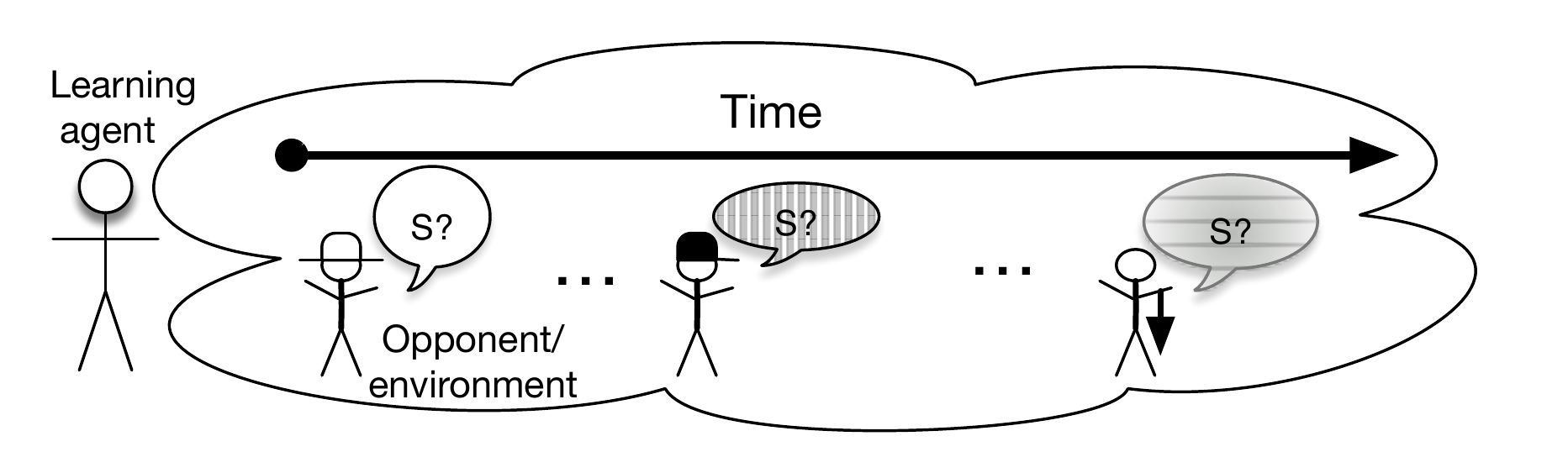}
\label{fig:categories:learn}
}
\vspace{-0.1cm}
\subfigure[Theory of mind: $\mathcal{O}$ reasons about how $\mathcal{A}$ might act and obtains a best response against that behaviour, \emph{BR($\mathcal{A}$}). $\mathcal{A}$ repeats that process with the model of $\mathcal{O}$, \emph{BR({BR($\mathcal{A}$}))}~\citep{Gmytrasiewicz:2000tx}. ] 
{
\includegraphics[scale=0.40]{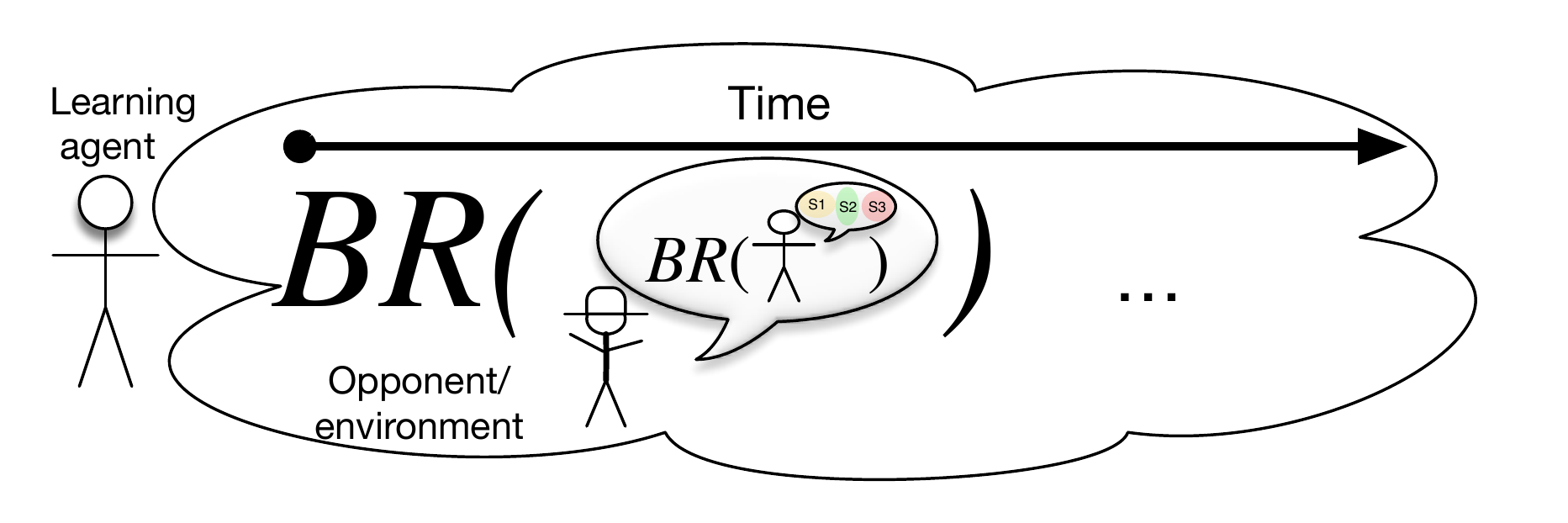}
\label{fig:categories:tim}
}
\caption{\small A learning agent $\mathcal{A}$ (outside the cloud) and how it models one opponent $\mathcal{O}$  (inside the cloud) exemplifying the 5 categories of how to handle non-stationary behaviour.}
\label{fig:categories}
\end{center}
\end{figure*}

In this section we exemplify each category and contrast them by using the same domain (see Figure~\ref{fig:categories}). Our example is presented in the context of the iterated prisoner's dilemma (see Section~\ref{sec:prelim:rg}) where two agents $\mathcal{A}$ and $\mathcal{O}$ play the infinite-horizon version of this game. 
\begin{definition}
Prisoner's dilemma (PD) is a normal-form game $ \langle \mathcal{N},A,u \rangle$, where:
	\item The set of players $\mathcal{N} = \{\mathcal{A}, \mathcal{O}\}$ are the two agents.
 	\item The set of actions is the same for both agents, they have two possible actions $A = \{C,D\}$.
 	\item$ u= (u_\mathcal{A}, u_\mathcal{O})$ where $u_i: A  \mapsto \mathbb R$ the payoff function for player $i$ as shown in Table~\ref{tab:prisionersdilemma}, satisfying: $t_{pd} > r_{pd} > p_{pd} > s_{pd}$  and $2r_{pd} > p_{pd}+s_{pd}$. 
\end{definition}
In the PD game when both players cooperate they both obtain the reward $r_{pd}$. If both defect, they get a punishment reward $p_{pd}$. If a player chooses to cooperate with someone who defects receives the sucker's payoff $s_{pd}$, whereas the defecting player gains the temptation to defect, $t_{pd}$. 

We now present slight variations of the above scenario exemplifying the assumptions made by algorithms in each category, pointing out where they are most useful and where their main assumptions do not hold.

\subsection{Ignore}
In this category algorithms can be useful with simple opponents or by making probably unrealistic assumptions, \emph{ignoring the non-stationary behaviour}. For example, assume the opponent uses a mixed (stationary) strategy, $\pi_m=(0.25,0.75)$ with higher probability of selecting \emph{defect}. If the assumption is correct, the learning agent can use fictitious play~\citep{Brown:1951vc} to learn an optimal policy against $\mathcal{O}$. However, consider the case that after $\mathcal{A}$ has learned the optimal policy, $\mathcal{O}$ decides to change to a Tit-for-Tat strategy $\pi_{TFT}$, thus $\mathcal{A}$'s learned policy will no longer be optimal.

\subsection{Forget}
Now, consider a different set of assumptions where $\mathcal{A}$ is interested in converging to a stationary policy and $\mathcal{O}$ has the same interest. Thus, both agents need to adapt to the changing (non-stationary) behaviour of the other (see Figure~\ref{fig:categories:forget}). One algorithm that is especially useful in this scenario is WoLF-PHC~\citep{Bowling:2002vva}, the algorithm generalizes \qlearning, but it was proposed to converge to a stationary policy in self-play. We can view WOLF-PHC as \emph{continuously learning (and forgetting), adjusting its learning rate to cope with the changing behaviour of the opponent}. Note that, if we remove the assumption of self-play (and we assume $\mathcal{O}$ uses a different behaviour), then WOLF-PHC loses its convergence guarantees.

\subsection{Respond to target opponents}
If the learning agent \emph{knows (or assumes) the opponent will behave in specific ways, that information can be used to target classes of opponents}. For example, assume $\mathcal{A}$ knows that the opponent will use the set of strategies \{\emph{Tit-for-Tat, Pavlov, Bully}\} and change among them stochastically. In this case HM-MDPs~\citep{Choi:1999tw} can target that type of opponent since they assume the environment can be represented in different stationary modes (MDPs) with stochastic transitions among themselves. Note that there are different algorithms that target a variety of classes (see Section~\ref{sec:algos:target}). However, if the assumptions about the opponent do not hold (in this case, adding a new strategy to the initial set) these algorithms provide restricted adaptability and therefore the policy will be suboptimal after most opponent changes.

\subsection{Learn opponent models}
In this category, agent $\mathcal{A}$ \emph{learns a model of the opponent which is used to derive an optimal acting policy. In this case, the learning agent starts without predefined opponent strategies or policies}~\citep{DaSilva:2006uw,HernandezLeal:2013dq}. Instead, $\mathcal{A}$ assumes the opponent will use several stationary strategies with infrequent changes among them. For example, in the iPD the opponent could start with \emph{Pavlov} and later change to \emph{Tit-for-Tat}. Moreover, if the opponent returns to a previous learned strategy, $\mathcal{A}$ should be able to detect and change its policy without relearning the model~\citep{HernandezLeal:2016uh}. However, one limitation of these algorithms is that they do not consider the strategic behaviour of $\mathcal{O}$ (an opponent that reasons about the agent $\mathcal{A}$).

\subsection{Theory of mind}
In the last category, the learning agent \emph{assumes an opponent that is performing strategic reasoning}. This is, in the lowest level $\mathcal{O}$ reasons about $\mathcal{A}$, in an upper level $\mathcal{A}$ reasons about $\mathcal{O}$ reasoning about $\mathcal{A}$. Best responding to a reasoning level is the way to obtain an acting policy. For example, assume the opponent thinks $\mathcal{A}$ uses a set of strategies to act \{\emph{Bully, random, Pavlov}\}, a distribution of those strategies represent the  zero level or reasoning, $L_{0}$. With the previous information $\mathcal{O}$ can compute a best response (\emph{BR}) against $L_{0}$, called level 1 strategy, $L_{1}=BR(L_0)$. Moreover, $\mathcal{A}$ can compute a best response against a distribution of the previous two levels, to obtain an acting policy (level 2) $L_{2}= BR(\{L_{1},L_{0}\})$. Note that this recursive reasoning could continue upwards and is the base of many approaches~\citep{Camerer:2004va,Gmytrasiewicz:2005un,Wunder:2010uv,Wunder:2012wj}. A limitation is that the basic strategies need to be specified \emph{a priori} and computing optimal policies can be computationally expensive~\citep{Gmytrasiewicz:2005un}.

In the next section we present an extensive list of state-of-the-art algorithms in from game theory, multi-armed bandits and RL and where they fall into each category of sophistication along with their environment and opponent characteristics.

% !TEX root = survey.tex

\section{Algorithms}
\label{sec:learningnonstationary}

\begin{table}
\centering
\tiny
\caption{\small A categorisation of different algorithms in terms of how they handle non-stationarity and with respect to related characteristics such as observability, opponent adaptation (described in Section \ref{sec:categories}) and the domain they were designed for: one-shot games (OSG), repeated games (RG), stochastic games (SG), extensive-form games (EG), sequential decision tasks (SDT) and multi-armed bandit scenarios (MAB).}
\begin{center}
\begin{tabular}{@{}p{6.0cm}lP{2.5cm}P{2.0cm}@{}} \toprule
\bf Category  &   &  &  \\ 
\bf \hfill Algorithm &  \bf Observability & \bf Opp. adaptation & \bf Designed for  \\ \hline
\bf  Ignore \\
\hfill Fictitious play~\citep{Brown:1951vc} & O. actions & No & RG   \\
\hfill \qlearning~\citep{Watkins:1989uk} & Local reward & No & SDT \\
\hfill JAL~\citep{Claus:1998tb}& O. actions  & No &  RG \\
\hfill UCB~ \citep{Auer:2002fd} & Local reward & No  & MAB\\
\hfill Exp3 and Exp4~\citep{Auer:2002vx} & Local reward & No & MAB \\
\hfill \rmax~\citep{Brafman:2003vk} &  O. actions and payoffs & No & SG \\ \hline
\bf Forget \\ 
%\hfill Satisficing~\citep{stimpson2001satisficing}  & Local rewards  & Slow & RG \\ 
\hfill WOLF-IGA~\citep{Bowling:2002vva}  &O. actions and rewards & Slow & SG \\
\hfill WOLF-PHC~\citep{Bowling:2002vva}  &Local rewards & Slow & SG \\
\hfill GIGA-WOLF~\citep{Bowling:2005vi}& Local rewards & Slow & RG\\ 
\hfill COLF~\citep{MunozdeCote:2006du}  &O. actions & Slow & RG\\
\hfill WPL~\citep{Abdallah:2008uua} & Local reward & Slow  & RG \\
\hfill WMD-UCB~\citep{Yu:2009tt} & Local rewards & Drastic & MAB \\
\hfill D-UCB~\citep{Garivier:2011br}& Local rewards  & Drastic & MAB \\
\hfill SW-UCB~\citep{Garivier:2011br}& Local rewards & Drastic  & MAB \\
\hfill FAQL / IQ~\citep{Kaisers:2010vx}& O. actions & Slow  & RG\\
\hfill \textcolor{black}{LFAQ~\citep{Bloembergen:2010wd}}& O. actions & Slow  & RG\\
\hfill \textcolor{black}{Rexp3~\citep{Besbes:2014uv}} & Local rewards &  Both & MAB\\
\hfill FAL-SG~\citep{Elidrisi:2014ux}&O. actions & Drastic & SG \\
\hfill \rmaxsharp~\citep{HernandezLeal:2016vv} & O. actions & Drastic & RG \\
\hfill RUQ~\citep{Abdallah:2013wfa,Abdallah:2016wn} & O. actions &  Slow & RG 
\\
\hfill \textcolor{black}{UUB~\citep{Lakkaraju:2017to}} & Local rewards &  Slow & MAB\\ \hline
\bf Target \\
\hfill Minimax-Q~\citep{Littman:1994ta} & O. actions & Drastic &  SG\\
\hfill Nash-Q~\citep{Hu:1998vu} & O. actions and payoffs & Slow &  SG\\
\hfill HM-MDPs~\citep{Choi:1999tw}   &O. actions &  Drastic & SDT \\
\hfill FF-Q~\cite{Littman:2001vc} & O. actions and payoffs & Slow &  SG\\
\hfill EXORL~\citep{Suematsu:2002cs} & O. actions and payoffs & Slow & SG \\
\hfill Exp3.S~\citep{Auer:2002vr} & Local rewards & Drastic & MAB \\
\hfill Hyper-Q~\citep{Tesauro:2003wq}  &O. actions & Slow & SG \\ 
\hfill Correlated-Q~\citep{Greenwald:2003uy}  & O. actions and payoffs & Slow &SG \\
\hfill NSCP~\citep{Weinberg:2004wj} &O. actions & Slow & SG\\  
\hfill ReDVaLeR~\citep{Banerjee:2004ve}& O. actions & Slow & RG\\  
\hfill MetaStrategy~\citep{Powers:2004wd} & O. actions and payoffs & 
Drastic & RG  \\
\hfill Manipulator~\citep{Powers:2005ws,Powers:2007gq} & O. actions and payoffs & Drastic  & RG\\
\hfill AWESOME~\citep{Conitzer:2006du} & Local rewards  & Slow  & RG \\
\hfill RNR and DBR~\citep{Johanson:2007ts,Johanson:2009uv}&O. actions  & Slow  & EG \\
\hfill ORDP~\citep{Yu:2009tv}  &O. actions  &  Drastic  & SDT \\
\hfill M-Qubed~\citep{Crandall:2011dt}  &O. actions  & Slow & RG \\  
\hfill \textcolor{black}{Pepper~\citep{Crandall:2012vo}}  &O. actions  & Slow & SG \\  
\hfill \emph{MDP-A} and BPR~\citep{Mahmud:2013wja,Rosman:2015vh}& O. actions & Drastic & SDT\\ 
\hfill \textcolor{black}{HS3MDPs~\citep{Hadoux:2014us}}   &O. actions &  Drastic & SDT \\
\hfill \textcolor{black}{RSRS~\citep{Damer:2017us}}&O. actions  & Slow  & RG \\
\hfill \textcolor{black}{OLSI~\citep{HernandezLeal:2017vi}} &O. actions  & Drastic & SG \\ 
\hline
\bf Learn \\
\hfill RL-CD~\citep{DaSilva:2006uw,Hadoux:2014tx} &O. actions  &  Drastic & SDT \\
\hfill $\zeta-$R-MAX~\citep{Lopes:2012ti}  &  O. actions & Slow & SDT  \\
\hfill CMLeS~\citep{Chakraborty:2013ii} & O. actions & Slow & RG\\
\hfill MDP-CL~\citep{HernandezLeal:2013dq}&O. actions &  Drastic & RG \\
\hfill Restless Markov bandits~\citep{Ortner:2014gu}&Local rewards &  Drastic & MAB \\
\hfill DriftER~\citep{HernandezLeal:2016twa} &O. actions  &  Drastic & RG\\ 
\hfill BPR+~\citep{HernandezLeal:2016uh,HernandezLeal:2016tw} &O. actions  & Drastic & RG \\ 
\hline
\bf Theory of mind  \\
\hfill RMM~\citep{Gmytrasiewicz:2000tx} & O. actions and payoffs & No & SDT \\ 
\hfill s-EWA~\citep{Camerer:2002ka}& O. actions and payoffs & Slow & RG \\ 
\hfill Level-K~\citep{CostaGomes:2001wt}& O. actions  &  No & OSG \\
\hfill Cognitive Hierarchy~\citep{Camerer:2004va}& O. actions  &  No & OSG \\
\hfill I-POMDP~\citep{Gmytrasiewicz:2005un}& O. actions &  No  & SDT \\
\hfill PI-POMDP~\citep{Wunder:2011vd,Wunder:2012wj}& O. actions & No & RG \\
\hfill \textcolor{black}{ToM/MToM~\citep{deWeerd:2013cg,VanderOsten:2017ty}}& O. actions & Slow & RG/SG \\
\bottomrule
\end{tabular}
\end{center}
\label{tab:stateoftheart}
\end{table}

\begin{figure}
\begin{center}
\includegraphics[scale=0.40]{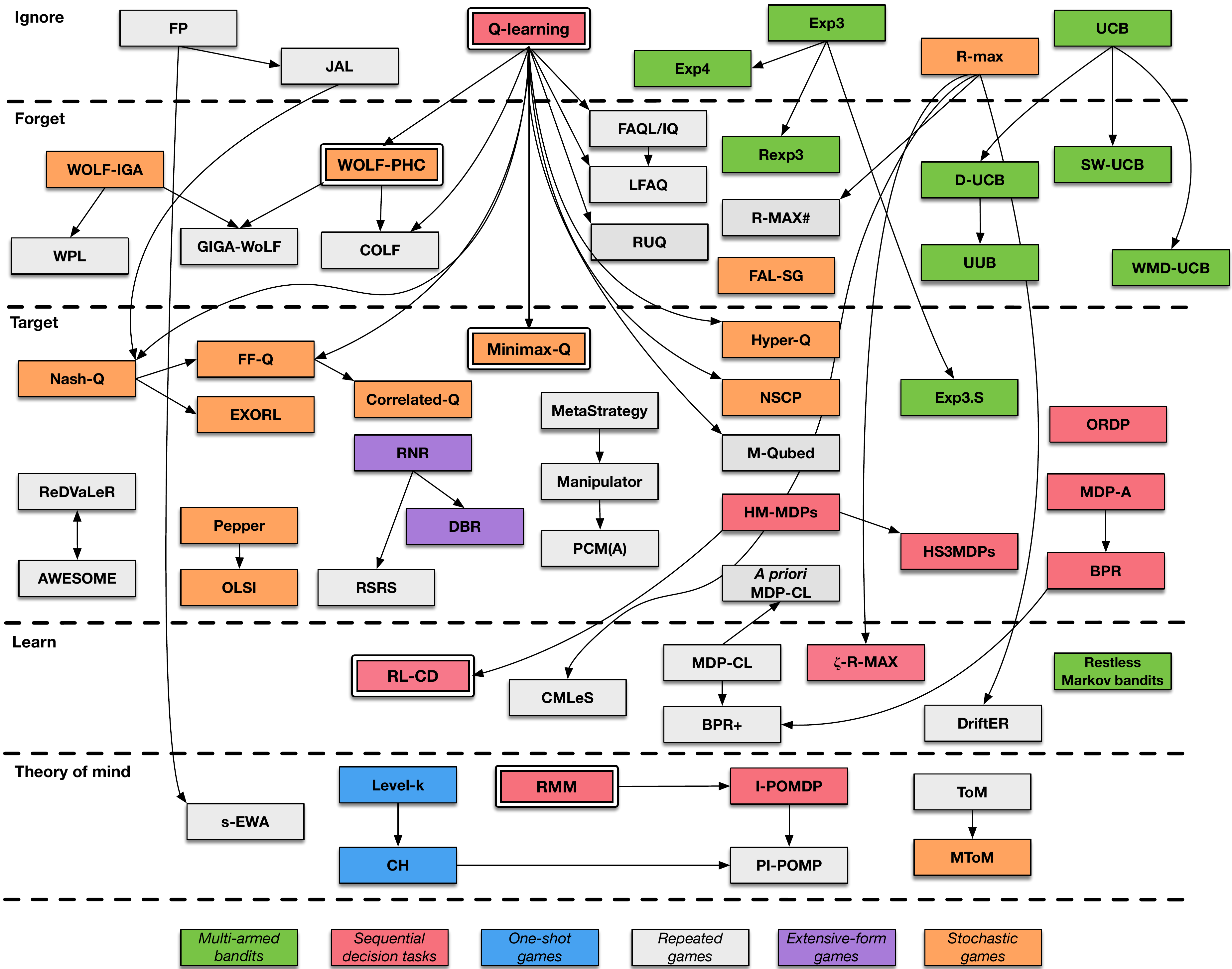}
\caption{\small %\revise{add a pointer between Q-Learning and JAL, Satisficing algorithm is missing} 
Diagram of the algorithms (coloured boxes; each colour represent one experimental domain) analysed in this survey divided in 5 categories (dashed lines) on how they handle non-stationarity. We present how they are connected to each other (arrows) and highlight those algorithms that are representative of each category (double box).}
\label{fig:relatedWork}
\end{center}
\end{figure}

In this section we present an extensive list of algorithms categorised with respect to how they deal with non-stationarity. Table~\ref{tab:stateoftheart} summarises this section by providing for each algorithm its category and some related characteristics such as observability, opponent adaptation and the environment it was designed for. Similarly, Figure~\ref{fig:relatedWork} depicts a diagram highlighting the connections among the algorithms and showing the most representative ones of each category.

%\reviseq{Section 5. Ignore is informally subdivided into game theory, bandits, RL; Forget is loosely subdivided in MAB, RL, and game theory; Respond to target opponents is divided in model-free approaches and model-based approaches. Since MAB, RL and game theory the three background sections, maybe keeping the same subdivision of algorithms in the subsections of section 5 will add some more structure to the paper?  This will especially be useful for those people looking for algorithms from either of those three fields specifically.}

\subsection{Ignore}
\label{sec:algos:ignore}

\textcolor{black}{Game theory is the study of strategic interactions among several agents, with the central concept of equilibrium among players denoting a mutual best response. While such reasoning does account for opponent strategies, classic algorithms typically do not account for \emph{changes} in opponent strategies. One early work for learning in repeated games is \textbf{fictitious play}~\citep{Brown:1951vc}. The model maintains a count of the plays by the opponent in the past. The opponent is assumed to be playing a \emph{stationary mixed strategy} and the observed frequencies are taken to represent the opponent's mixed strategy. However, if the opponent does not follow a stationary strategy the method will not compute a best response. %Also there are several games where it does not converge to the Nash equilibrium~\citep{Shapley:1964ut}.
}

Techniques from multi-armed bandits have been used to deal with the exploration-exploitation trade-off. In the classical bandit setting some assumptions are made regarding the rewards, e.g., the rewards are drawn independently from \emph{some fixed (stationary)}, but unknown distributions. In this context, the \textbf{UCB (upper confidence bounds)} algorithm~\citep{Auer:2002fd} guarantees low regret under certain conditions. UCB uses the principle of optimism in the face of uncertainty to select its actions (assumes an optimistic guess on the expected rewards). A different setting is the adversarial bandit setting where no statistical assumptions are made about the generation of rewards \citep{Auer:2002vx}. Instead, the reward associated with each arm at every round are \emph{fixed in advance} by an adversary before the game starts. Even in this complicated scenario, the \textbf{exponential-weight algorithm for exploration and exploitation (Exp3)} provides theoretical bounds for expected rewards~\citep{Auer:2002vx}. Finally, a different algorithm based on Exp3 is the \textbf{exponential-weight algorithm for exploration and exploitation using expert advice (Exp4)}~\citep{Auer:2002vx}. The scenario is different since now the algorithm assumes a set of ``experts'' that provide a mechanism to select an action. Exp4 provides bounds of expected utility to perform nearly as well as the best expert in hindsight. Note that these bandit algorithms assume the setting is \emph{fixed in advance} and does not account for changes during the interaction (i.e., \emph{ignore}).

In the context of RL, a model-based algorithm for acting optimally in adversarial environments is \textbf{\rmax}~\citep{Brafman:2003vk}. The algorithm uses an MDP to model the environment which is initialised optimistically assuming all actions return the maximum possible reward, (r-max). After several experiences with the environment \rmax updates \emph{and fixes} a part of the model (i.e., state-action pairs). The policy efficiently leads the agent to less known state-action pairs or exploits known ones with high utility. \rmax promotes an efficient sample complexity of exploration~\citep{Kakade:2003vv}, this means that \rmax has theoretical guarantees for obtaining near-optimal expected rewards. However, \rmax alone will not work when the environment presents \emph{non-stationary} behaviour~\citep{Lopes:2012ti} since it fails to adjust its model if the environment changes.  
 
The classic model-free RL algorithm of \textbf{\qlearning} assumes a \emph{stationary} environment. However, it has been applied with some success in different multiagent scenarios~\citep{Tan:1993em,Sen:1994ud,Crites:1998ui}. The simple approach of using plain \qlearning, i.e., \emph{ignoring} other agents in the environment, is known as independent learners. In contrast, \textbf{joint-action learners (JALs)}  model the strategies of the opponents explicitly by taking into account the joint-action of all the agents in the \qlearning update, which implies the agent can observe the actions of others~\citep{Claus:1998tb}. However, even when they have more information, convergence is not dramatically enhanced. Moreover, in JALs it is required a considerable amount of contrary experience to be overcome some changing behaviour~\citep{Claus:1998tb}.
 
% \reviseq{I would say Q-Learning and JAL can fall under Ignore or Forget, depending on their learning parameter setting; if alpha = 1/k, then it’s ignore; if alpha is a fixed number, then it’s forget, since older information will gradually be lost, and new information will predominate}
\subsection{Forget}
\label{sec:algos:forget}

Failing to update with current information is the main limitation of the algorithms in the previous category. A solution is to \emph{forget} old information and update with recent one, which has been experimentally noted to improve learning algorithms in repeated games~\citep{Bouzy:2010wr}. While this category could be more aptly described as \emph{adaptive discounting of experiences}, we chose the label \emph{forget} as an intuitive concept and mnemonic.

Even though UCB has been empirically shown to not work well on non-stationary environments~\citep{Hartland:2006wo}, it has inspired two algorithms that can adapt to sudden changes, i.e., when the distributions of rewards changes abruptly (not depending on the policy of the player or on the sequence of rewards). \cite{Garivier:2011br} proposed two methods: the \textbf{discounted UCB (D-UCB)} whose policy averages past rewards with a discount factor giving more weight to recent observations; and the \textbf{sliding-window approach (SW-UCB)} which relies on a local empirical information of the last $\tau$ plays ($\tau$ being a parameter of the algorithm). \textcolor{black}{ \cite{Lakkaraju:2017to} proposed a variant of D-UCB to solve an exploration problem where the expected utility of each arm is non-stationary. However, \emph{instead of assuming arbitrary changes in the utility distribution (as D-UCB), their setting has certain structure} which is encoded in their proposed \textbf{bandit for unknown unknowns (UUB)} algorithm.} \cite{Yu:2009tt} tackle a specific non-stationary bandit problem with two main characteristics: (i) the rewards are piecewise-stationary, i.e., the reward distribution changes arbitrarily and at arbitrary time instants, but it remains stationary on intervals; (ii) the agent can observe some of the past outcomes of arms that have not been picked. \cite{Yu:2009tt} propose the \textbf{windowed mean-shift detection (WMD)-UCB} to cope with these scenarios. The algorithm works by detecting changes in the environment using a statistical test on the most recent $\tau$ time-steps (i.e., sliding window), when this happens the algorithm resets. Note that discounting or using a sliding window approach have the same effect, give more weight to recent observations and \emph{forgetting} the old one.

\textcolor{black}{A different way to model non-stationarity in multi-armed bandit scenarios is to \emph{assume the total variation in expected reward is bounded by a (known) \emph{variation budget}}. This allows to model diverse reward changes, e.g., both slow and continuous or drastic jumps. \cite{Besbes:2014uv} proposed the \textbf{Rexp3} algorithm (based on Exp3) for this setting and their results highlight a trade-off that exists between retaining and forgetting information, i.e., the fewer past observations to recall, the larger the associated error; the more past observations, the higher the chances of these being biased towards outdated information.}

In the context of efficiently exploring adversarial environments one example of the forgetting behaviour is the \textbf{\rmaxsharp} algorithm~\citep{HernandezLeal:2016vv}. \rmaxsharp proposes a drift exploration to detect changes that happened in the opponent, but that may have not been noticed, which results in suboptimal behaviour. This effect is known as \emph{shadowing}~\citep{Fulda:2006tu} or observationally equivalent models~\citep{Doshi:2006tv}. To avoid this effect, the  solution is to continually revisit states that have not been visited recently (which is determined by a parameter). Therefore \rmaxsharp proposes to reset (to \emph{r-max}) those state-action pairs and then update the model and policy which will implicitly re-explore those parts of the environment. \rmaxsharp provides theoretical results showing that under some assumptions it is guaranteed to learn a new model within finite sample complexity. Note that, in contrast to the classic \rmax which fixes one part of its model and later is never allowed to update that same part; \rmaxsharp is continually updating its model (and policy) to keep up with the non-stationary environment. However, the approach may not be easily scalable to scenarios with many agents.

In the model-free context of RL there are two variants of \qlearning that achieve convergence of self-play in specific games by updating action-value estimators equally fast, even when one action is more frequently selected than another: The first has been studied under the name \textbf{frequency-adjusted \qlearning (FAQL)}~\citep{Kaisers:2010vx,Kaisers:2011uf} and individual \qlearning (IQ)~\citep{Leslie:2005}, and the second one is \textbf{repeated update \qlearning (RUQ)}~\citep{Abdallah:2013wfa,Abdallah:2016wn}. 
Intuitively, the action that receives fewer updates needs to make larger adjustments to keep up, which is implemented with a learning rate modulation in FAQL/IQ (inversely proportional to the probability of the action's selection probability), and by repeated updates in RUQ. As a result, all actions receive the same expected learning speed. Formally, these learning speed modulations make it possible to prove the limit behaviour of the algorithms in self-play converges to Nash distributions in zero-sum games~\citep{Leslie:2005}, with convergence points shown to approach Nash equilibria as the exploration temperature decreases in two-agent two-action games~\citep{Kaisers:2011uf,Kianercy:2012hf}.
\textcolor{black}{\cite{Bloembergen:2010wd} proposed \textbf{lenient frequency adjusted \qlearning (LFAQ)} for cooperative multi-agent environments. This extension incorporates the concept of \emph{leniency}~\citep{Panait:2006tx} to account for initial mis-coordination, which enables LFAQ to obtain high convergence to Pareto optimal equilibria in cooperative games.}
Note that convergence results of this type of algorithms require the assumption of infinite interactions and/or infinitesimal learning rates. 
Effectively, the action-value estimates of frequently selected actions are expected to be more recent and accurate, receiving only small updates based on each new observation. In contrast, scarcely selected actions are likely to have older action-value estimates, which in non-stationary environments may become less accurate with age, and therefore more weight is put into the new observation--the value estimator is updated with a larger learning rate towards the new observation. As a consequence, these algorithms can be said to implement a dynamic strategy to \emph{forget outdated action-value estimates}.

The win or learn fast (WoLF) principle was introduced to make an algorithm that (i) converges to a stationary policy in multiagent systems and (ii) if other players' policies converge to stationary policies then the algorithm should converge to a best response \citep{Bowling:2002vva}. The intuition of WoLF is to learn quickly when losing and cautiously when winning. One proposed algorithm that uses this principle is \textbf{WoLF-IGA (infinitesimal gradient ascent)}. The algorithm at each interaction updates its strategy (in the direction of the gradient) to increase its expected payoffs with some fixed step size. WoLF-IGA has been proved theoretically to converge in self-play in a two-person, two-action repeated matrix games. However, WoLF-IGA assumes to know an equilibrium from the start which can be complicated in many games.  \textbf{Generalized IGA (GIGA)-WoLF}~\citep{Bowling:2005vi} improves on WoLF-IGA in two aspects. First, it does not need to known an equilibrium strategy. Second, it also addresses the challenge of not being exploited by an opponent by showing no-regret in the limit~\citep{Bowling:2005vi}. Finally, another practical variant of the WoLF principle is \textbf{WoLF policy-hill climbing (WoLF-PHC)}~\citep{Bowling:2002vva}, which is based on \qlearning and performs hill-climbing in the space of mixed policies. To cope with \emph{non-stationary} behaviour WoLF-PHC changes between two learning rates depending on how the algorithm sees the interaction is happening, i.e., by comparing whether the current expected value is greater than the current expected value of the average policy. 

%\revise{A \textbf{satisficing learning} strategy was proposed by \cite{stimpson2001satisficing}. The concept of satisficing is a modification of the rationality concept from game theory; rather than calculating the optimal actions, the agent defines a \emph{satisfaction} level, when the level is met it continues like that, otherwise explores for better actions. Note that the aspiration level can be adaptive.}

In the context of cooperative game theory it is common to look for Pareto efficient solutions. On one side, recall that WoLF algorithms aim converge to the Nash equilibrium, thus they are not the best candidate for this different type of problems~\citep{Goodrich:2003wu}. On the other side, using simple \qlearning algorithms results in suboptimal solutions due to the parallel learning process which makes the environment non-stationary. To overcome this issue, \textbf{CoLF (change of learn fast)}~\citep{MunozdeCote:2006du} is another algorithm inspired by the WoLF principle, but with the objective of promoting cooperation of self-interested agents to achieve a Pareto efficient solution in repeated games. CoLF proposes to adjust the learning rate of the algorithm depending on the received rewards: slow when unexpected or changing (i.e., \emph{non-stationary}) and fast when they are stable, \emph{near-stationary}. Note that changing the learning rates is a common method to keep up with non-stationary environments. In the end this adaptation results in updating information and \emph{forgetting outdated estimates}.

\textbf{Weighted policy learner (WPL)}~\citep{Abdallah:2008uua} is another algorithm designed to converge to a Nash equilibrium. However, in contrast to previous algorithms it can do so with limited knowledge observing only local rewards (the agent neither knows the underlying game nor observes other agents actions). WPL share some similarities with WoLF-IGA since it also has two modes for adjusting its learning rate, however there are also some key differences: (i) WPL needs considerably less information and (ii) WPL uses a continuous spectrum of learning rates (WOLF-IGA uses two fixed ones).

%One approach that falls outside of the RL paradigm is the
\textbf{Fast adaptive learner (FAL)}~\citep{MohamedElidrisi:2012wm} is designed to learn quickly in two-player repeated games. The algorithm is based on two components: (i) to predict the next action of the opponent the entropy learning pruned hypothesis space (ELPH) algorithm is used, ELPH is an online learning algorithm that maintains a set of hypotheses according to a \emph{fixed window of the history of observations}~\citep{Jensen:2005ui}. The frequency count of each hypothesis is used to obtain the entropy which is used as an indicator of the quality of the prediction. (ii) To obtain a strategy against the opponent the authors use a modified version of the Godfather strategy.\footnote{The Godfather strategy gives the opponent the opportunity to cooperate with an action that is beneficial for both players. If the opponent does not accept the offer, Godfather will force the opponent to obtain its security level~\citep{Littman:2001vaa}.} An extension of FAL for stochastic games is FAL-SG \citep{Elidrisi:2014ux}. To deal with this different setting, FAL-SG abstracts the stochastic game into a meta-game matrix via clustering, after which the original FAL approach can be used.

\subsection{Respond to target opponents}
\label{sec:algos:target}

Previous approaches updated their behaviour according to the newest information available, in contrast, algorithms in this group have a pre-defined target of opponents. This is the category with the largest number of algorithms. The reason is that easier to provide guarantees against specific opponents than against general classes; to better understand the different approaches we made subdivision for this category into model-free and model-based approaches.

%\vspace{0.5cm}
\paragraph{Model-free approaches}\mbox{} \\

\noindent In the context of multi-armed bandits one extension of Exp3 is \textbf{Exp3.S}~\citep{Auer:2002vr} which \emph{targets} a  specific adversarial bandit scenario in which the \emph{bandits are allowed to shift $S$ times} (a parameter of the algorithm). The algorithm keeps track of the alternative which gives highest reward even if this best alternative changes over time. The algorithm guarantees low regret assuming the number of shifts ($S$) and the number of rounds in the interaction is known in advance. 

In the traditional single-agent version of \qlearning the objective is to maximise the sum of rewards in an environment. In contrast, \textbf{Minimax-Q} proposes to extend \qlearning to zero-sum stochastic games, \emph{assuming an opponent which has a diametrically opposed objective} to the agent. The algorithm uses the minimax operator to take into account the opponent actions~\citep{Littman:1994ta}. This allows the agent to converge to a fixed strategy that is guaranteed to be safe in that it does as well as possible against the worst possible opponent (the agent tries to maximize its rewards and the opponent aims to minimise those). The algorithm is guaranteed to converge in self-play to a stationary policy. Nevertheless, there are cases when minimax-Q does not converge to the best response, i.e., is not rational~\citep{Bowling:2002vva}. 

\textbf{Hyper-Q}~\citep{Tesauro:2003wq} is another extension of \qlearning designed for multiagent systems (specifically for stochastic games). The main difference that the $Q$ function depends on three parameters: the state, \emph{the estimated joint mixed strategy of all other agents}, and the current mixed strategy of the agent. Hyper-Q assumes that only the opponents' actions (not the payoffs) are observable. To obtain an approximation of the mixed strategies a discretisation has to be performed and the $Q$-table could easily grow exponentially in the number of discretisation points. Hyper-Q is guaranteed to converge to the optimal value function against the following three groups of opponents: (i) stationary opponents, (ii) non-stationary opponents that define its history-independent strategy depending only on themselves and not on the Hyper-Q player (e.g., replicator dynamics model, see \citealp{Borgers:1997jw}) and (iii) non-stationary opponents that accurately estimate the Hyper-Q agent strategy and then adapt using a fixed history-independent rule.

\textbf{M-Qubed (Max or Minimax \qlearning)}~\citep{Crandall:2011dt} is a RL algorithm designed for two-player repeated games. The authors mention several compromises which an algorithm needs to balance: bounding loses (safety), playing optimally (best respond) and taking risks for ensuring cooperation and coordination. To achieve this, the algorithm \emph{targets two groups of opponents} and proposes \emph{different behaviours (best-response and cautious) against each group}. M-Qubed typically selects actions based on its Q-values updated via SARSA (best-response), but triggers to a minimax strategy when its total loss exceeds a pre-determine threshold (cautious).

Another targeted set of opponents consist of \emph{agents using non-stationary policies with a limit} (i.e., decreasing possibly infinite changes). The \textbf{non-stationary converging policies (NSCP)} algorithm~\citep{Weinberg:2004wj} it is based on \qlearning and computes a best response to opponents in which the probability that the strategy would be far away from the limit gets smaller as the rounds increase. For this, \cite{Weinberg:2004wj} define a distance between two stage game strategies as the distance between the probability vectors of the strategies. An example of this type of opponent is start with a uniform distribution over a set of actions and at each time-step the probability slowly moves towards one action with probability 1 and the rest with 0.

Previous algorithms aim to best-respond to target opponents, however, another common approach is to \emph{respond with the aim of converging to a Nash equilibrium}. \textbf{Nash-Q}~\citep{Hu:1998vu,Hu:2003wba} is a variation of \qlearning that needs to observe the opponent actions and rewards to converge in some cases. The algorithm update Q-values over joint actions rather than a single-agent Q function. Another main difference with respect to \qlearning is that it updates with future payoffs \emph{assuming all agents will use a NE strategy}. \textbf{Friend-or-foe \qlearning (FF-Q)}~\citep{Littman:2001vc} generalise Nash-Q and Minimax-Q algorithms. FF-Q \emph{treat each opponent either as friend or foe} and can converge in two cases: adversarial (minimax) equilibrium or in coordination games with unique equilibrium. Furthermore, a generalization of FF-Q is \textbf{Correlated \qlearning}~\citep{Greenwald:2003uy} which instead of converging to a Nash equilibrium, it looks for a correlated equilibrium\footnote{In these games it is assumed a public signal from the environment which is observed by all agents, a real-world example is a traffic signal, the agents decide its strategy based on that signal.}  which is more general than a NE~\citep{Aumann:1974im}. A common problem regarding NE is the selection when there are multiple options, to deal with this issue Correlated-Q uses four equilibrium selection functions which depending on the objective to maximise (e.g., each individual reward, the sum of the players rewards). However, to compute any of those it needs to observe opponents' actions and rewards.

A limitation of previous approaches is that they target only one group (class) of opponents. Therefore, some algorithms improve on that regard, one example is \textbf{EXORL (extended optimal response)}~\citep{Suematsu:2002cs} which has \emph{two main acting behaviours: best response or Nash equilibrium}. EXORL starts learning a best response to the opponent (using on-policy learning), but if the opponent adapts (determined by a parameter) then it will look for a Nash equilibrium. 
\textbf{Replicator dynamics with a variable learning rate (ReDVaLeR)}~\citep{Banerjee:2004ve} builds on the same ideas of EXORL: best response against stationary opponents and NE against adaptive opponents. Moreover, ReDVaLeR adds another characteristic, constant bounded expected regret at any time against any number of opponents~\citep{Banerjee:2004ve}. This makes the algorithm more robust since it is \emph{implicitly targeting opponents that are neither stationary nor using the same learning algorithm}. ReDVaLeR needs to observe opponent actions, if this is not possible then \textbf{AWESOME (adapt when everybody is stationary otherwise move to equilibrium)}~\citep{Conitzer:2006du} is designed for this case. AWESOME converges to a Nash-equilibrium in self-play and when the opponents seem stationary it will learn a best response and can do so with limited information (i.e., only local rewards). 

We noted that algorithms basically target three main behaviours depending on the opponents: convergence (against adaptive opponents), best response (against stationary opponents) and bound the loss (against other types of opponents). In this regard, \cite{Powers:2004wd} formalised these three properties as compatibility, targeted optimality and safety. Moreover, they proposed the \textbf{MetaStrategy}~\citep{Powers:2004wd} algorithm that achieves those three properties by alternate among the strategies: fictitious play, minimax and a modified Bully.\footnote{\cite{Littman:2001vaa} proposed the Bully strategy which is an example of a Stackelberg leader.} A slightly different algorithm is \textbf{Manipulator}\footnote{PCM(A) is an extension of Manipulator to multiplayer games~\citep{Powers:2007gq}.} \citep{Powers:2005ws} which alternates among: best response, minimax and a modified Godfather strategy. Moreover, Manipulator has the same guarantees as MetaStrategy against a richer class of target opponents, \emph{memory-bounded opponents}.\footnote{In the same context of bounded memory adversaries, but in the bandit setting~\cite{Arora:2012ta} showed that no bandit algorithm can guarantee a sublinear policy regret against an adaptive adversary with unbounded memory. However, if the adversary's memory is bounded, they propose a technique which converts any bandit algorithm with sublinear regret bound into a sublinear policy regret bound.} These are defined as opponents that play a conditional strategy where actions can only depend on recent periods, this is its distribution over actions can only depend on the most recent $k$ periods of past history.
We note that the approach of how MetaStrategy and Manipulator decide on which strategy to use is the same: (i) first to explore (ii) to determine how the opponent reacts and  possibly act with a best response; (iii) otherwise the algorithms opt for a safe option (minimax strategy).

\paragraph{Model-based approaches}\mbox{} \\

\begin{figure}
\center
\includegraphics[scale=0.4]{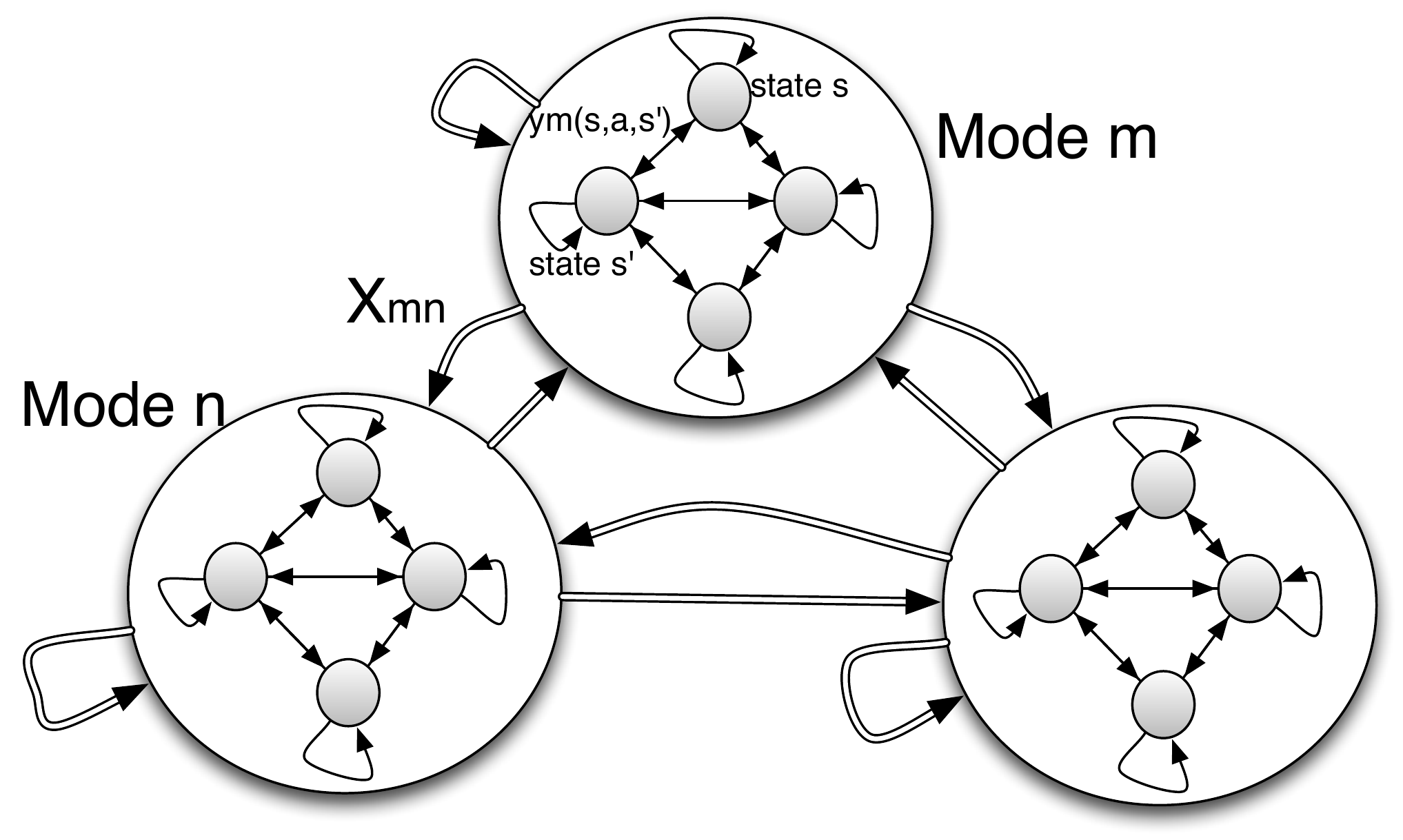}
\caption{\small An example of an HM-MDP with 3 modes (large circles) and 4 states (smaller shaded circles). The value $X{mn}$ represents a transition probability between modes $m$ and $n$, and $ym(s,a,s')$ represents a state transition probability in mode $m$.}
\label{fig:hmmdpExample}
\end{figure}

\noindent \textbf{Hidden-mode Markov decision processes (HM-MDPs)} are a model-based technique to deal with non-stationary environments~\citep{Choi:1999tw}. They assume \emph{the environment can be represented in a small number of modes}. Each mode is a stationary environment, which has different dynamics and needs a different policy. It is assumed that at each time-step there is only one active mode. The modes are hidden, which means that cannot be directly observed, they are only estimated by past observations. Moreover, transitions between modes are stochastic events. Each mode is modelled as an MDP. Different MDPs along with its transition probabilities form an HM-MDP which can be seen as a special case of a POMDP~\citep{Choi:2001tg}. Figure~\ref{fig:hmmdpExample} depicts an example of an HM-MDP with 3 modes and 4 states. Each of the three large circles represent a mode, shaded circles inside the modes represent states. Thick arrows indicate stochastic transitions between modes and thinner arrows represent state-action-next state probabilities. A limitation of HM-MDPs is that they need to fix the number of modes from the start and do not provide any form of online learning.

\textcolor{black}{
HM-MDPs assume the environment may change at every timestep, which may not hold in many environments. \textbf{Hidden-semi-Markov-mode Markov decision processes (HS3MDPs)}~\citep{Hadoux:2014us} take inspiration from HM-MDPs to represent problems in non-stationary environments but their difference is that HS3MSPs assume these changes evolve according to a semi-Markov chain (i.e., when the environment stochastically changes to a new environment it stays in that environment during a stochastically drawn duration). HS3MDPs are equivalent to HM-MDPs and form a subclass of POMDPs. To solve large-sized HS3MDPs, Hadoux et al. proposed an adaptation of POMCP~\citep{Silver:2010ws}.
}

Learning an opponent model is usually a way to obtain an acting policy (see Section~\ref{sec:algos:learn}). However, some algorithms \emph{assume to start with a set of policies to act} and the problem now becomes which policy to select in an non-stationary environment. For example, \cite{Mahmud:2013wja} propose a scenario in which a latent variable changes rarely, but when it happens it modifies the optimal policy. Thus, the agent at the beginning of each round selects one policy from a known (and predefined) set $\Pi$. The goal of the agent is thus to select policies to minimise the total regret incurred in the limited task duration with respect to the performance of the best alternative from $\Pi$ in hindsight. \textbf{\emph{MDP-A}}~\citep{Mahmud:2013wja} was designed for single agent scenarios with a set of tasks in which an agent needs to perform similar tasks (the same state and action space), but with different policies for each task. MDP-A uses a transfer learning approach in which given a collection of source behaviour policies, eliminates the policies that do not apply in the new task using a statistical test in an online fashion. Similarly, \textbf{Bayesian policy reuse (BPR)} \citep{Rosman:2015vh} is another approach that draws inspiration from \emph{MDP-A} since they work under the same scenario. However,  BPR computes a belief distribution over the tasks and with every step of interaction it receives a signal which is used to update that belief using the Bayes rule. A limitation of BPR is the \emph{assumption of knowing a priori ``performance models''} (probability distributions) describing how policies behave on different tasks. 

A similar problem has been studied in the context of repeated games. \cite{HernandezLeal:2014wo} analysed a scenario where \emph{the opponent has a set of stationary strategies and changes among them during the interaction}. Moreover, they assumed to know those strategies (represented as MDPs, see~\citealp{Banerjee:2005wq}) before the interaction. \textbf{\emph{A priori} MDP-CL} is an algorithm designed to quickly detect the strategy used by a non-stationary opponent~\citep{HernandezLeal:2014wo}. \emph{A priori} MDP-CL explores with different actions for a period of rounds to learn an opponent model  in the form of an MDP which is compared to the initially known strategies. If the learned model matches one of the prior known opponent strategies then the exploration phase finishes and the agent can solve the MDP (that represent the opponent) to obtain an policy against it. 

\textcolor{black}{Inspired by the paradigm of optimism in face of uncertainty~\citep{Brafman:2003vk}, \cite{Crandall:2012vo} proposed the  \textbf{potential exploration with pseudo stationary restarts (Pepper)} algorithm to learn in repeated stochastic games.  Pepper creates a family of new algorithms when plugged together with learning algorithms for repeated matrix games (e.g., M-Qubed, see~\citealp{Crandall:2011dt}; fictitious play, see~\citealp{Brown:1951vc}). \cite{HernandezLeal:2017vi} proposed a variation of repeated stochastic games in which the opponent may change constantly and its identity is unknown to the learning agent. Their \textbf{opponent learning in sequential interactions (OLSI)} algorithm is generalization of Pepper to learn from different opponents by \emph{keeping a belief over a hypothesised opponent set}.}

Note that most approaches respond optimally only to targeted opponents, but remain silent to what happen against other opponents. Against this background, \cite{Johanson:2007ts} analysed how robust are counter strategies (learned to act against a single opponent) across different opponents in a poker domain. They performed this analysis using the modelling technique called frequentist best response (FBR). Their analysis showed that FBR is very successful at exploiting the opponent it was designed to exploit. However, when FBR strategies play against other opponents their performance is poor. To solve this issue, the authors propose the \textbf{restricted Nash response (RNR)} algorithm to generate robust strategies against a specific opponent, but at the same time they assume the opponent may slightly change~\citep{Johanson:2007ts}. The strategy obtained by RNR is based on defining a probability $p$ for the opponent to act as the learned model and  with probability $1-p$ it will act different than the model. RNR requires a large number of observations and sometimes can over fit the opponent model. Later,  \textbf{data biased response (DBR)}~\citep{Johanson:2009uv} which extends from RNR was proposed to overcome those problems.
\textcolor{black}{Recently, \textbf{restricted Stackelberg response with safety (RSRS)}~\citep{Damer:2017us} was proposed to find a \emph{robust} response against an opponent in normal-form games. As RNR, RSRS uses the confidence in its prediction over the opponent, however, RSRS adds a \emph{safety margin which reflects the level of risk it is willing to tolerate}, which results in a trade-off between best-responding to the prediction and providing a guarantee of worst-case performance.}

Another algorithm which put emphasis on robustness is the \textbf{online robust dynamic programming (ORDP)}. \cite{Yu:2009tv} presented ORDP for an extreme case of non-stationary behaviour; instead of assuming a adversary with a fixed objective ORDP \emph{assumes the opponent may play an arbitrary sequence of actions}. This translates into arbitrary variations in the reward function and arbitrary, but bounded, variations in the transition probabilities. Since solving this problem is computationally  expensive, ORDP has another (lazy) version which provides a trade-off between performance and computational complexity~\citep{Yu:2009tv}.

\subsection{Learn opponent models}
\label{sec:algos:learn}

Algorithms in the previous category share as deficiency that they target a specific opponent, but with limited adaptability if the opponent does not follow their assumptions. To cope with this problem algorithms in this category learn an opponent model and use it to derived an acting policy. Updating that model (and therefore the policy) is the way to keep up with against a non-stationary opponent.

Recall the scenario where \emph{the environment changes (infrequently) among several stationary modes and the agent needs to update its policy accordingly}. In this scenario, \cite{DaSilva:2006uw} proposed the the \textbf{reinforcement learning with context detection (RL-CD)} algorithm where the stationary environments are called \emph{contexts} for which a partial model is learned (for example, using Dyna-Q;~\citealp{suttonBarto1998}). At each time-step RL-CD decides which partial model to use according to a quality measure and when all partial models seem far then it starts learning a new model. \textcolor{black}{\cite{Hadoux:2014tx} proposed an adaptation of RL-CD, replacing the quality measure by statistical tests for change-point detection, yielding \textbf{RL-CD with sequential change-point detection}. In a similar vein, \cite{Banerjee:2016vu} proposed the \textbf{quickest change detection (QCD)} approach based on a two-threshold strategy to detect model changes in MDPs (with changes in transition and/or rewards).}
  
Previously we presented the adversarial bandit scenario (see Section~\ref{sec:algos:ignore}). Later, we presented algorithms that use either a \emph{forget mechanism} (see Section~\ref{sec:algos:forget}) or \emph{target a specific bandit switching scenario} (see Section~\ref{sec:algos:target}). Lastly, \textbf{restless Markov bandits}~\citep{Ortner:2014gu} are another specific bandit scenario in which the stochastic process governing each arm does not depend on the actions of the learner, instead \emph{it depends on a Markov chain which transitions independently whether the learner pulls that arm or not}. Note that in this case the problem becomes a partially observable setting \citep{Ortner:2014gu}. Also one main characteristic of the setting is that  the optimal policy cannot always be expressed in terms of arm indexes. \cite{Ortner:2014gu} proposed to treat this problem as learning an MDP, in particular they use a modification of the URCL2 algorithm~\citep{Jaksch:2010wb} for which they provide regret bounds.

In the context of efficient adversarial exploration, the \textbf{$\zeta$-\rmax} algorithm~\citep{Lopes:2012ti} extends from the classical \rmax. Recall that \rmax fixes a state-action pair after sufficient visitations. This has the drawback of not consider the actual empirical prediction performance or learning rate of the learner w.r.t. the data seen so far \citep{Lopes:2012ti}.
In contrast, $\zeta$-\rmax \emph{estimate the learning progress in terms of the loss over the training data used for model learning.} The idea is to compute a $\zeta$ function which is based on the leave-one-out cross validation error. $\zeta$-\rmax handles changes in the environment better than \rmax while also having a \emph{PAC-MDP} efficient guarantee. A limitation of this approach is the computational cost of computing $\zeta$, since it depends on the number of states and actions at every iteration.

Memory-bounded opponents have been of interest in the MAL community (see Section~\ref{sec:algos:target};~\citealp{Powers:2005ws,Powers:2007gq}). However, previous approaches dot not actively seek to learn an opponent model. In contrast, \cite{Banerjee:2005wq} proposed to \emph{learn a model of those opponents whose policy is a (fixed) function of some historical window of past joint-actions by all the agents.} The adversary induced MDP (AIM)~\citep{Banerjee:2005wq} is a technique for repeated games which induces an MDP that implicitly has modelled the opponent (stationary) strategy. The learning agent, by knowing the MDP that the opponent induces, can compute an optimal policy $\pi^*$. These types of players can be thought of as a finite automata that take the most recent actions of the opponent and use this history to compute their policy~\citep{MunozdeCote:2010wy}. These AIM models have been used as basis to derive other learning algorithms. One of those is the \textbf{convergence with model learning and safety (CMLeS)}~\citep{Chakraborty:2013ii}. CMLeS achieves three results: (i) convergence to following a Nash equilibrium joint-policy in self-play; (ii) targeted optimality (close to best response) against memory-bounded agents whose memory size is upper bounded by a known value; and (iii) safety (ensures an individual return that is very close to its security value).

Another approach that uses AIMs to model opponents is the \textbf{MDP-CL (continuous learning)} algorithm~\citep{HernandezLeal:2013dq}. The algorithm was proposed \emph{to act optimally against non-stationary opponents that switch among several stationary strategies}. MDP-CL starts without prior models or polices and uses an exploratory phase (random actions) for a determined number of rounds. After this phase, \emph{it computes a model of the opponent in the form of an MDP which yields an optimal policy}. In this point it starts learning another model (which will be used to detect changes) and after some rounds (defined by a parameter) the MDP-CL agent make comparisons between the learned models to evaluate their similarity. If the distance between models is greater than a given threshold, it is determined that the opponent has changed strategy and the modelling agent must restart the learning phase, resetting both models and starting from scratch with a random exploratory strategy. Otherwise, it means that the opponent has not switched strategies and the optimal policy is being used. \textbf{DriftER (drift based on error rate)}~\citep{HernandezLeal:2016twa} is another algorithm designed for \emph{acting against switching non-stationary opponents}. DriftER uses \rmax as exploratory policy instead of a random exploration and to detect switches it draws inspiration from concept drift~\citep{Widmer:1996tt}. DriftER uses the learned MDP to predict the opponent actions and to keep track of their model quality. Moreover, DriftER provides guarantees of switch detection with high probability~\citep{HernandezLeal:2016twa}. A limitation of both DriftER and MDP-CL is that they assume a period of rounds where the opponent will remain stationary in which the model learning take place. 

Finally, it is worth mentioning a scenario where \emph{a switching opponent either can use a new strategy (unknown to the other agent) or a return to a previously used one}, in this cases it will be useful only to learn the unknown strategy and quickly detect when it is a known strategy. \textbf{\bprnew}~\citep{HernandezLeal:2016uh,HernandezLeal:2016tw} which is extension of BPR~\citep{Rosman:2015vh} is designed for these scenarios. \bprnew assumes a non-stationary opponent that switches among stationary strategies. The algorithm starts without prior models or policies, therefore during the interaction it learns an opponent model and when the opponent changes (detected by low performance) it is stored it its memory which might be eventually useful if the opponent returns to that same strategy. 

\subsection{Theory of mind}
\label{sec:algos:tom}

Approaches in the previous category learned models of other agents in the environment in order to derive an acting policy. In this last category of sophistication we present algorithms that do not only model opponents' behaviour, but also assume a strategic reasoning about the opponent, which represents a nested (or recursive) reasoning.

In this category we distinguish algorithms which either are inspired by two main areas behavioural game theory and planning (see Section~\ref{sec:relatedModels}).
In the former category we found the \textbf{level-k} and \textbf{cognitive hierarchy} models which have been used to model human interactions~\citep{Camerer:2004va,Stahl:1995wl,CostaGomes:2001wt}. These are also known as \emph{iterative reasoning models}, which refers to approach they take to make decisions. The general concept involves an initial set of zero level strategies, this is without strategic behaviour (for example, randomizing across all actions). \emph{Computing a best response against the lower level forms the base of the next level}. However, most of these approaches have been studied only in the context of one-shot games. One exception is the work by \cite{Wunder:2010uv} in which they a model populations consisting of agents with different reasoning levels in the iterated prisoner's dilemma. The way to act optimally against the population was obtained by best responding using a cognitive hierarchy model \citep{Camerer:2004va} which was modelled as a POMDP~\citep{Littman:1996ts}.

\textbf{Sophisticated experience-weighted attraction (s-EWA)}~\citep{Camerer:2002ka} is another behavioural game theory algorithm inspired by fictitious play. It assumes two types of opponents, (simple) adaptive opponents (using the EWA, see~\citealp{Camerer:1999ua}) and \emph{sophisticated opponents that rationally best-responds to her forecasts of all other behaviours} (they use the s-EWA algorithm). A limitations is that it has been only studied in the context of short repeated games (less than 10 rounds).

In the planning category, one of the earliest approaches proposed by~\cite{Gmytrasiewicz:2000tx} is the \textbf{recursive modelling method (RMM)}. They propose \emph{a specialized knowledge representation in the form of reward matrices that allows using a recursive reasoning to obtain the best coordinated action in a MAS system}. An approach inspired in RMM, but with a formal decision theoretic background are the \textbf{interactive POMDPs (I-POMDPs)}~\citep{Gmytrasiewicz:2005un}. They are called interactive because the model considers what an agent knows and believes about what another agent knows and believes~\citep{Aumann:1999wb}. This means that \emph{an agent will have a model of how it believes another agent reasons}. I-POMDPs extend POMDPs incorporating models of other agents into the regular state space. The main limitation of these models is its inherent complexity, since solving one I-POMDP with $M$ number of models considered in each level, with $\ell$ maximum reasoning levels, is equivalent to solving  $O(M^\ell)$ POMDPs~\citep{Seuken:2008eh}. Despite these issues, there are recent algorithms for online learning~\citep{BrendaNg:2012wb}. Also there are works using I-POMDPs with more than a thousand of agents~\citep{Sonu:2015ts} and even in experiments with humans~\citep{Doshi:2010wl}. \textbf{Parametrized I-POMDPs (PI-POMDPs)}~\citep{Wunder:2011vd,Wunder:2012wj} are an approach which combines I-POMDPs with the iterative reasoning models. The idea is \emph{to compute a policy that maximizes the score against either a distribution over previous levels, or a selection of agents from those levels}, by solving the POMDP formed by them. While computationally expensive it provides a clear formalism to work showing good results in highly adaptive domains, such as the lemonade stand game \citep{Zinkevich:2011kb}. However, further work is needed to show the applicability to other domains.

\textcolor{black}{
Lastly, another theory of mind model was proposed by \cite{deWeerd:2013cg}. Here, the zero-level is composed of beliefs indicating the likelihood of the opponent taking any action at any state, higher order models are generated based on the information from lower levels. Additionally, they use \emph{a confidence value} which helps the agent to adapt to different opponents (with different levels of reasoning). Recently, an extension to more than one opponent, \textbf{multiagent ToM (MToM)}, was proposed by~\cite{VanderOsten:2017ty}. To cope with this challenge the authors propose a \emph{stereotyping mechanism} (clustering), which segments the agent population into sub-groups of agents with similar behaviour; these groups are then treated as single agents. 
}

We have presented the five categories of how algorithms deal with non-stationary and classified state of the art algorithms with different characteristics. The next section presents the strengths and limitations of each category, related areas to multiagent learning, and pinpoints open avenues for future research.

\section{Discussion}
\label{sec:discussion}

%\reviseq{Section 6. The discussion section feels like a bag of ‘other concepts’ thrown together and not like a discussion. The first section (6.1) is very valuable, as it lists the advantages and limitations of each approach and suggests when to use which approach; except in the case of the theory of mind, where  suddenly  neutral language is used and do not give a recommendation for when it should be used - adjust this please. This whole subsection can be longer as it should present the core discussion of the paper. The environments in section (6.2), are a bit out-of-place. It is useful to know which problems are often used, but should this be part of the ‘discussion’? The theoretical results section is another useful list of citations, but does not elaborate much on the cited works - not much discussion. In the related areas section,  the “multi-agent interaction without prior coordination” is not related, but should rather be part of the research topic of the paper and should therefore be discussed in section 5.4 or 5.5. The future research section is good.}

We have presented five categories of how learning algorithms deal with non-stationary behaviour. In this section we start by discussing their strengths and limitations (see Section~\ref{sec:strenghtsandlimitations}). Then, we mention the most common experimental domains that have been used (see Section~\ref{sec:experimentaldomains}), we outline the current theoretical results (see Section~\ref{sec:theoreticalresults}), and we describe related areas of research (see Section~\ref{sec:relatedModels}). We conclude with exploring promising avenues of future research (see Section~\ref{sec:openquestions}).

\subsection{Strengths and limitations of each category}
\label{sec:strenghtsandlimitations}

We briefly mention some advantages and limitations for each category and provide pointers for when each category is especially useful.

\begin{description}
\item [Ignore.]  These algorithms are widely known in the community and most of them do not need to known extra information from the opponent (opponent’s payoffs). However, a large drawback is that most of them lose their theoretical guarantees when used in non-stationary environments (e.g., \qlearning). We advise to use this algorithms where no extra information can be obtained from the environment.

\item [Forget.] One advantage of these algorithms, in contrast to the previous category, is that they do take into account the non-stationarity of a multiagent system.  In general, these algorithms are model-free approaches with the limitation that they might take longer periods to converge to a solution \citep{Suematsu:2002cs}. These algorithms could be used when no a priori information is known about the opponent and there are no constraints in the learning time.

\item [Respond to target opponents.] If the opponent is restricted to a single class (i.e., worst case opponent, see~\citealp{Littman:1994ta}; stochastically changing among models, see~\citealp{Choi:1999tw}; converging to a Nash equilibrium, see~\citealp{Hu:1998vu}) then algorithms in this category offer an efficient solution. A limitation is the constrained adaptability of these algorithms. For example, if we expect the opponent to use a wider set of strategies then the solution is to directly learn a model of the opponent.

\item [Learn opponent models.] A main advantage of these algorithms is that the learned model of the opponent can be reused if the opponent returns to the same strategy~\citep{DaSilva:2006uw,HernandezLeal:2016uh}. Since these algorithms are model-based, they usually learn faster than other approaches. One limitation is that they need the opponent to remain stationary for a long enough period to model them, which can be unrealistic in some scenarios. 

\item [Theory of mind.] An interesting feature of algorithms in this category is that they are readily available to model populations (more than 2 agents) since that is the intrinsic way they obtain an acting policy~\citep{Camerer:2004va, Wunder:2010uv, Wunder:2012wj}. Another characteristic of these algorithms is that they perform a complex strategic reasoning process, which necessitates high computational costs to solve them (e.g., I-POMDPs, see~\citealp{Seuken:2008eh}). Also, these approaches have been studied mostly for predicting behaviour in unrepeated games~\citep{Wright:2014dm}.%\todo{What does the final sentence convey here? How does it relate to an advantage/disadvantage?}
\end{description}

\subsection{Experimental domains and applications}
\label{sec:experimentaldomains}

Most testing scenarios for multiagent interactions use the formal models of game theory, from extensive-form games, repeated games to stochastic games. However, there is also another category of specific applications, such as negotiation, smart grids, and routing problems.

\paragraph{Extensive-form games}
The classic game of poker has different variations ranging from simple to complex (in terms of the state space and action space) which have been used to evaluate different opponents. Kuhn poker is a tiny, toy variant of poker. The game involves two players, two actions and a three card deck. This game has been studied previously since the two players strategies can be summarized in two or three parameters~\citep{Hoehn:2005vk,Bard:2007wg}. Leduc hold'em Poker is a larger version than Khun Poker in which the deck consists of six cards~\citep{Bard:2015ta}. Heads-up limit Texas hold'em is more complex variation, where the game tree consists of approximately $9.17 \times 10^{17}$ states~\citep{Johanson:2007ts}. Given the size of the domain, algorithms have focused on dealing with this problem~\citep{Bard:2013ta}. %\todo{This suggests that the environment is not non-stationary in the Kuhn poker variant, but is that what we mean?}
%\reviseq{I don’t understand why the extensive-form games paragraph should be all about poker, it sounds like it’s the only extensive-form game.}

\paragraph{Repeated games}
It is common to use repeated games as a setting with non-stationary opponents~\citep{Suematsu:2002cs,Bowling:2002vva,Tesauro:2003wq,Weinberg:2004wj,Powers:2005ws,Conitzer:2006du,Abdallah:2008uua,Crandall:2011dt,MohamedElidrisi:2012wm,HernandezLeal:2013vw,HernandezLeal:2013dq,HernandezLeal:2016uh,Damer:2017us}. The most simple games have two players and two actions (2x2); a 3x3 example is rock-scissors-paper. Also, it is common to evaluate learning algorithms in randomly generated games according to certain specifications such as zero-sum games or, constant-sum games \citep{Nudelman:2004vq}. Previous works have performed experimental comparisons among different multiagent learning algorithms in repeated games~\citep{Bouzy:2010wr}.

One interesting competition which can be represented as a repeated game is the lemonade stand game~\citep{Zinkevich:2011kb}. Here, three agents (vendors) interact by choosing a position (12 different actions) on an ``island'' in order to sell lemonade to the island's population. The rewards depend on the actions of all the agents and several interesting algorithms were developed in this context where fast adaptation was needed~\citep{Wunder:2010wc,MunozdeCote:2010wq,Sykulski:2010vo,Wunder:2011vd}.

\paragraph{Stochastic games}

This type of games generally poses a more difficult challenge than repeated games since there are different states (games) with probabilistic transitions (see Section~\ref{sec:prelim:gametheory}). Many stochastic games represent grid-worlds, where agents need to take strategic decisions. For example, a mini-version of the sports game  \emph{soccer} was proposed as  a stochastic game played on a 4x5 grid with five actions and two players, an attacker and the goal keeper~\citep{Littman:1994ta}. %\revise{Add a citation and make the dictintion with the real game} 
In this game, agents must use a probabilistic policy to obtain higher rewards~\citep{Littman:1994ta,Bowling:2002vva}. Other interesting games are stochastic versions of well-known games such as PD, coordination, and chicken~\citep{MunozdeCote:2008ua,Elidrisi:2014ux}. %\todo{any reason to \emph{italicize} soccer, but not the other games?}

\paragraph{Other domains}

%\reviseq{- Section 6.2 includes a section on “other domains.” It seems to me that each of the problems mentioned in this subsection could be modeled as a stochastic game of some kind of formalized game alreadydiscussed. Why the distinction? It appears that this domain just includes scenarios that the authors do not want to formalize.}

Lastly, many algorithms have been evaluated in specific applications ranging from aerospace to security and surveillance; for a complete survey about the impact of MAS applications refer to~\cite{Muller:2014uo}.

A typical situation where non-stationary multi-agent learning plays an important role is automated negotiation and e-negotiation systems~\citep{Jen01,Kra01}. Recent examples include the setting of e-commerce~\citep{He03,Kow10}, virtual agents~\citep{Devault2015,Gratch2015} and games such as diplomacy~\citep{Fabregues:2010} and coloured trails~\citep{Gal:2005,Lin:2010}. As with human negotiations, automated negotiation between agents is a non-stationary game with incomplete information, where the agents initially do not know their opponent’s preferences and where strategies can change over time (for a survey on learning in negotiation, see~\citealp{BaarslagOMSurvey}). As a result, they need to derive information from the exchange of offers with each other. 

Although rarely framed in the context of non-stationary learning, many automated negotiation strategies have been formulated that take advantage of non-stationary learning mechanisms. An important category is preference learning, in which agents aim to learn aspects of the opponent's preference profile by engaging in online opponent \emph{model learning} in an effort to reach Pareto optimal (win-win) outcomes (e.g.,~\citealp{Coe04,Hin09Benefits,BaarslagOpponentModels,Zha15}). Learning the opponent's negotiation strategy is another important aspect, which boils down to determining counter-offers in subsequent negotiation states. The agents face the challenge of a wide diversity of possible negotiation strategies and the fact that the opponent can change behaviour dynamically according to the offers received~\citep{Hou04,Baarslag11PRIMA}. That is, learning the opponent's strategy is a \emph{moving target} problem, where the agent simultaneously seeks to acquire new knowledge about the opponent while the agent needs to optimize its negotiation actions based on the current model. In the negotiation literature, \emph{responding to target opponents} is a opponent model classification problem, where the type of the opponent needs to be determined from a range of possibilities given its negotiation behaviour~\citep{Lin08}. There also exist simple \emph{ignore} and \emph{forget} strategies that either assume a stationary environment or only employ recent data, for example negotiation tactics that take into account elapsed time only~\citep{Far98}. More recently, automated negotiators have even been endowed with (second-order) \emph{theory of mind}, so that agents can reason about what the opponent believes about their beliefs~\citep{deWeerd:2015,Pynadath:2013}. An important negotiation domain involves smart energy grids and their trading markets used to buy and sell energy. The Power TAC simulator~\citep{Ketter:2013kn} models a complex a dynamic energy system in this context, where different brokers can take actions in three markets. One of those is the wholesale market, which is a particular type of auction. The non-stationary behaviour appears when there are brokers that switch among different strategies through time~\citep{HernandezLeal:2015tq}. Another example is the problem of predicting the energy demand of users, which involves randomness and changes in behaviour  \citep{Marinescu:2014hf,Marinescu:2015wt}. 

Routing problems have been also treated as a domain with non-stationary behaviour. In domain routing, an ISP operator has the opportunity to increase its revenue by charging external domains for the traffic transiting on its links. Moreover, agents must be able to deal with a non-stationary environment when the optimal price setting varies according to other ISPs' strategies and the network load~\citep{Vrancx:2015dn}. In the context of smart-cities, there are different routing problems that model non-stationary behaviour, such as traffic networks. In this case, the world is represented as a grid, with traffic lights on each junction and patterns of traffic representing different stationary environments~\citep{Choi:1999tw,DaSilva:2006uw}. 

This section presented experimental domains commonly used in non-stationary environments while the next section focuses on theoretical results.

\subsection{Theoretical results}
\label{sec:theoreticalresults}
In this section we outline different theoretical results presented in the context of learning in non-stationary environments.

\paragraph{Regret bounds.} Multi-armed bandits algorithms, usually provide regret bounds for different algorithms and different types of scenarios (adversarial, stochastic, Markov chain; see~\citealp{Auer:2002fd,Auer:2002vr,Auer:2002vx,Yu:2009tt,Garivier:2011br,Ortner:2014gu,Besbes:2014uv}). Few algorithms provide regret bounds for sequential decision problems~\citep{Yu:2009tv} or multiagent scenarios~\citep{Bowling:2005vi}. %\revise{p34  In the regret bounds paragraph, do all the cited papers’ bounds hold under non-stationarity? I don’t think it is the case for the Auer et al 2002a paper. Correct me if I’m wrong.}

\paragraph{Efficient exploration guarantees.} Another category of theoretical results comprises those algorithms which provide efficient exploration guarantees (for example, using sample complexity results; see~\citealp{Kakade:2003vv}) in adversarial stationary environments~\citep{Brafman:2003vk} and non-stationary ones~\citep{Lopes:2012ti,HernandezLeal:2016vv}. 

\paragraph{Convergence to Nash equilibrium.}
A large group of algorithms has provided guarantees to converge to a NE under slightly different conditions:
only local rewards~\citep{Abdallah:2008uua},  partial observations~\citep{Conitzer:2006du}, complete information settings~\citep{Bowling:2002vva,Hu:1998vu,Littman:2001vc,Suematsu:2002cs}. Most of these algorithms assume NE only in self-play~\citep{Hu:1998vu,Bowling:2002vva,Banerjee:2004ve,Chakraborty:2013ii} or variations of self-play~\citep{Bowling:2005vi}.
 
\paragraph{Best response.}
\qlearning loses its guarantees (convergence to an optimal policy) in non-stationary environments. Because of that, most algorithms try to improve on that regard. For example, by still having guarantees in stationary environments, but also better suited for non-stationary environments~\citep{Abdallah:2016wn}. Other address directly non-stationary opponents and prove that will learn a best response policy~\citep{Tesauro:2003wq,Weinberg:2004wj,Chakraborty:2013ii}. 

\paragraph{Robustness guarantees.}
Another common result is to assess the robustness of an algorithms by providing guarantees of safety, security or no-exploitability in the form of expected rewards~\citep{Littman:1994ta,Johanson:2007ts,Johanson:2009uv,Powers:2007gq,Crandall:2011dt,Chakraborty:2013ii,Elidrisi:2014ux,Damer:2017us} or regret bounds~\citep{Yu:2009tv,Besbes:2014uv}. A different class of results is to provide switch detection guarantees against non-stationary opponents~\citep{HernandezLeal:2016twa} which makes the method robust. 

Next, we introduce different areas and paradigms that share a connection with learning in the presence of non-stationary behaviour. Later, we present open avenues of future research.

\subsection{Related areas}
\label{sec:relatedModels}
This section presents concepts and areas that might be useful to take into consideration when developing new algorithms.

\paragraph{Supervised learning and concept drift.}
 \begin{figure}
 \centering
 \includegraphics[scale=0.60]{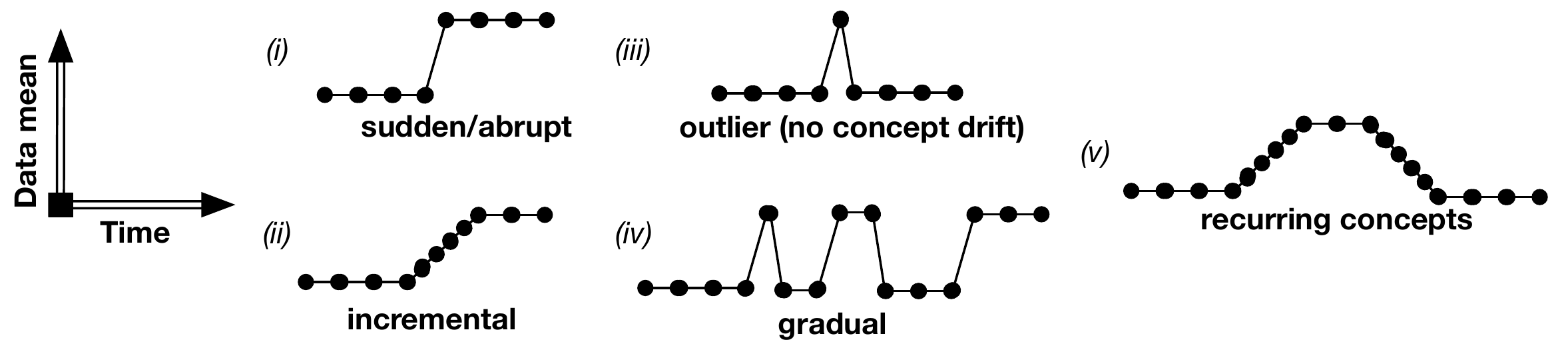}
 \caption{\small Types of concept drift that change over time \citep{Gama:2014cr}: (i) sudden, (ii) incremental, (iii) outliers, (iv) gradual and (v) recurrent.}
 \label{fig:conceptDriftTypes}
 \end{figure}

The machine learning community has developed an area related to non-stationary environments and online learning which is called \emph{concept drift}~\citep{Widmer:1996tt}. The approach is similar to a supervised learning scenario where the relation between the input data and the target variable changes over time. \cite{Gama:2014cr} presented an survey of this problem where different types of concept drift where categorized as depicted in Figure~\ref{fig:conceptDriftTypes} (using a one-dimensional data where changes happen in the data mean). (i) A change may happen suddenly/abruptly (from one time-step to the next). (ii) Incrementally, where there is a window of time where intermediate concepts appear. (iii) Outliers or noise, which refers to random deviation or anomaly, in which case no adaptation is needed. (iv) Gradually, where the concepts alternate one to another until finally converging to a different one. (v) Recurring, where previously seen concepts may reappear after some time. Concept drift scenarios are related to non-stationary environments, however they need to be adapted to a multiagent setting where there is a need for exploration in the form of action selection and uncertainty due to opponent's actions. However, some work in multiagent learning have drawn inspiration from concept drift \citep{HernandezLeal:2016twa}. 

%\paragraph{Change point detection.}
%Change-point detection objective is to find abrupt changes in time-series \citep{Yamada:2013ui}. There are several approaches in this area, most of them focused on large datasets \citep{KIFER:2004kv,Yamada:2013ui}. Bayesian approaches have also been proposed to calculate the probability of the current run (time since the last change-point) \citep{Adams:2007ti}. 
%Explaining the changes is another current interesting area where supervised learning algorithms have been proposed for interpreting the changes \citep{Ide:2008vo}. 

\paragraph{Transfer learning.} 
RL has been shown successful in many domains when a single agent is performing a single task (with the appropriate learning time). However, when having different tasks the basic approach is to learn a completely new model. To reduce this time consuming process, \emph{transfer learning} algorithms use the experience gained in learning to perform one task to improve learning performance in a related, but different, task~\citep{Taylor:2009ur}. This is especially important in some types of non-stationary environments. For example, in case of recurring changes (see Figure~\ref{fig:conceptDriftTypes}), previous information (for example, in the form of models or policies) will be useful to quickly have an acting policy. \textcolor{black}{These ideas (e.g., reusing past policies) have inspired recent works on multiagent systems~\citep{HernandezLeal:2016uh,HernandezLeal:2017vt}.}

\paragraph{Multiagent interaction without prior coordination.} 
\cite{Stone:2010wi} presented the challenge of \emph{ad-hoc teamworks}, this is, to create an autonomous agent that is able to efficiently and robustly collaborate with previously unknown teammates on tasks to which they are all individually capable of contributing as team members. Similarly, \emph{ad-hoc coordination} is the problem of designing an agent that is able to be flexible and efficient in a multiagent system that admits no prior coordination among the agents~\citep{Albrecht:2013tea}. \textcolor{black}{This active line of research~\citep{Barrett:2014vp,Melo:2016ic,Albrecht:2016is,Liemhetcharat:2017cx,Chakraborty:2017wx} is related since the agents involved can be of different types (heterogeneous agents) and they can have different adaptation behaviours which posses a problem since prior coordination is restricted.}

\paragraph{Partial observability and planning}
MDPs are the main model used by RL algorithms. However, there are other related models which are particularly relevant to the multiagent community. When cooperative teams of agents are planning in uncertain domains, they must coordinate to maximise their (joint) team value, in this scenario the \emph{multiagent Markov decision processes} (MMDPs)~\citep{Boutilier:1996tl} are useful. This model is a $n-$person stochastic game where the payoff function is the same for all agents. Currently there is undergoing research for reducing the costs related to computing these models \citep{Scharpff:2016te}. POMDPs, \emph{partially observable MDPs}~\citep{KaelblingLP:1998vs} are models where it is no longer the case that the agent has full perception capabilities. Instead, there is probability distribution over observations. In this way, it is possible to model problems in a more realistic way, the downside is that solving a POMDP is computationally more expensive than an MDP~\citep{Papadimitriou:1987wb}. Note that a particular case of a POMDP is the HM-MDP (see Section~\ref{sec:algos:target}). A generalization of POMDPs to a multiagent scenario with cooperative agents (since they need to share they utility function) are \emph{decentralized POMDPs}~\citep{Seuken:2008eh}. One limitation is its complexity which is NEXP-complete~\citep{Seuken:2008eh}. Recent works have proposed different methods to overcome this limitation, for example by searching in the \emph{influence space} (i.e., the space that represents probabilistic effects that agent policies may exert on one another, see \citealp{Witwicki:2012wf,Oliehoek:2015vs}).

\paragraph{Evolutionary game theory.} The application of game theoretic reasoning to the study of populations, initially to understand biological processes such as evolution, has received its own designation as \emph{evolutionary game theory}~\citep{weibull1995}. Initial work bringing this field towards multi-agent learning algorithms has established the formal link between the simple reinforcement learning algorithm \emph{cross learning} and the \emph{replicator dynamics}, a central concept in evolutionary game theory~\citep{Borgers:1997jw}. This has inspired a stream of follow-up work that links stochastic multi-agent reinforcement learning algorithms to varieties of deterministic dynamical systems, as summarized in a related survey~\citep{Bloembergen:2015ei}. The principle methodology is taking the limit of infinitesimal learning rates, and studying the resulting dynamical system to gain insight into the emergent behaviour of the multi-agent system, such as its convergence, stability and resilience. Additional interest is given to each equilibrium's basin of attraction and resulting welfare, providing an assessment of the anticipated joint interaction outcome. %expected outcomes based on an a priori distribution over initial states.

\paragraph{Behavioural game theory.} 
Many models proposed from a game theoretic approach do not accurately predict human-behaviour in many experiments~\citep{Kahneman:1979wl, Goeree:2001vy}. New models that take into account human characteristics (e.g., fairness, reciprocity, deception) were grouped under the name of \emph{behavioural game theory}~\citep{Camerer:2004va}. Even when these models tend to obtain good results in one-shot games with human populations \citep{Wright:2010vd} these models are still not well studied in repeated games or sequential decisions problems. 

Having mentioned closely related areas, we now present some interesting avenues for future research.

\subsection{Open questions and promising avenues of future research}
\label{sec:openquestions}
Although learning in multiagent systems has been an active research area in the past years there are still many open questions. In this section, we present four promising lines of research and we give example research questions that fall within each line. 

In a previous survey, \cite{Tuyls:2012up} presented three main challenges in MAS. We pinpoint some connections between those challenges and our proposed lines of research. In particular, for the ``extending the scope of MAL'' challenge, we propose ideas in the context of \emph{diversity in opponents} (see Line 1), \emph{dynamic interactions} (see Line 2) and \emph{applications} (see Line 4). Similarly, for the ``classification limitations'' challenge (a lack of classifications of what is missing in MAL), we proposed two ideas related to \emph{learning objectives} (see Line 3).
% \comment{this needs to be refined}.
%\todo{Should the note on the connections with Tuyls work be so prominently placed here?}

\paragraph{Line 1: Diversity in opponents}
\begin{itemize}
\item{Heterogeneous learning agents.} %Theory of mind algorithms provide a solid foundation for modelling strategic opponents. 
In real settings, one might encounter several agents with different learning characteristics, objectives, actuators, and representation of the world (including sensors). This heterogeneity is one of the most (if not the most) important complicating factors in acting optimally. One way to cope with such rich and complex environments is to characterise them (i.e., the set of learning opponents) across different labels, like diversity (i.e., how many types of learning agents), type distribution (i.e., the density distribution function) and set of learning techniques; e.g., if the learning agents are mostly using regret minimization or reinforcement learning or Bayesian non-parametrics, to name a few examples. This is an important strategy that has been used in the negotiation  literature where agents can establish optimal bidding strategies against specific types of opponents encountered in the environment~\citep{Mat98,ANAC2011}. In this way, one can constrain solutions to some well-defined subset of multiagent environments. \textcolor{black}{We encourage new algorithms to frame their work in the context of our proposed framework (see Section~\ref{sec:newFramework}) which naturally accounts for \emph{heterogeneous} opponents.}

\item{Modelling populations.} There are many complications when interacting with many agents, and for this reason, most algorithms use few agents in the environment. However, using those same algorithms could become intractable in large multiagent domains. To obtain efficient and scalable algorithms one would need to sacrifice detail by generalising the system to a population level, in a way to best respond to classes of populations rather than individuals~\citep{Wunder:2011vd,Bard:2015ta,HernandezLeal:2017vi,VanderOsten:2017ty}. \textcolor{black}{A different approach is to determine the degree of interaction among agents, this could help in defining whether to interact with an agent or ignore it and take it as part of the environment~\citep{DeHauwere:2010vq,Yu:2015cm}.} %%Encounters and related ideas

\item{Unknown world knowledge by opponents.} Algorithms in the learn category assume the agent is aware of the knowledge of all opponents, i.e., attributes or features that correctly describe the opponents' observations of the world. However, in most real situations this information is not really accessible~\citep{Chakraborty:2013vg}. To relax this assumption, the agent needs to learn the model and at the same time the correct knowledge representation~\citep{Maillard:2013tr}. A possibility to learn without putting effort into designing the correct representation is to use deep learning techniques~\citep{Deng:2013vh,Mnih:2015jp}. Another option to dealing with uncertain world representations by the opponents is to keep a set of known representations, as in \citep{HernandezLeal:2016uh}, and infer the correct one by maintaining (and updating) a probability distribution over the set. \textcolor{black}{This can be naturally modelled in the proposed framework (see Section~\ref{sec:newFramework}) where \emph{beliefs} over \emph{opponent behaviors} are two main components.}
\end{itemize}

\paragraph{Line 2: Dynamic interactions}
\begin{itemize}

\item{Learning in multiple concurrent interactions.}
Many multiagent learning algorithms assume interactions occur synchronously and among all agents. However, in real-world scenarios this is not always the case where interactions are usually asynchronous with different agents taking different response times. This holds especially true in large multi-agent coordination and negotiation systems where multiple, concurrent threads have to be coordinated. Communication protocols for committing and decommitting to deals have only been studied recently~\citep{Ito:2009,Wil:2012}. It is still an open questions whether current learning algorithms will work under these slightly different conditions.

\item{Intelligent reuse of information to reduce learning times.} 
Learning a model of the other agents in the environment is a way to solve the non-stationarity problem. However, this learning process usually requires a large period of repeated interactions, which is unreasonable in many scenarios. To alleviate this problem, information from previous interactions can be reused. For example, by generating a ``portfolio" of the possible opponents (in an offline phase) and during the interaction estimate which is the most similar and act with a respective policy~\citep{Bard:2013ta,Barrett:2013uh,Albrecht:2016jm,HernandezLeal:2016uh}\textcolor{black}{, please refer to our proposed framework which naturally models this approach (see Section~\ref{sec:newFramework}). Similarly, areas derived from transfer learning~\citep{Taylor:2009ur} could be extrapolated to multiagent scenarios such as \emph{curriculum learning}~\citep{Svetlik:2016wx,Narvekar:2017wc} where existing techniques work for a single agents (independently) and therefore an open question is to reuse information from different agents. Giving advice to agents~\citep{Torrey:2013wv,Zhan:2016vj} is another example in which multiagent algorithms are still in an early stage of development~\citep{daSilva:2017wy}.}

\item{Interaction against a dynamic number of opponents.}
In multiagent systems, the number of agents in the environment is usually fixed before the interaction and remains constant during the interaction. However, it is possible to consider the opponents may come and go during the interaction (e.g., dynamic coalition formation) which will affect the environment and most probably the acting policy. One idea on how to model these type of scenarios is to model each (dis)appearance of the opponents as a switch in the environment and use algorithms designed for these cases~\citep{DaSilva:2006uw,HernandezLeal:2016uh,HernandezLeal:2016twa}.

\item{Exploratory learning noise.}
The assumption to not explicitly model other agents and just consider them as part of the environment presents different problems. One of those happens when many learning agents explore at the same time creating noise to the rest, this is called exploratory action noise~\citep{HolmesParker:2014th,MunozdeCote:2006du}. \textcolor{black}{To alleviate this problem different methods have been proposed~\citep{Tumer:2007vs,HolmesParker:2014th}. Recently, the same problem appeared in a deep multiagent RL setting~\citep{Foerster:2017uq} and the proposed solution was based on previous work in the area~\citep{Tumer:2007vs}.} However, this problem is more complicated to solve in scenarios where no coordination or cooperation is possible. 
\end{itemize}

\paragraph{Line 3: Learning objectives }
\begin{itemize}

\item{Tracking vs convergent algorithms --- transient performance.}
One way to categorise learning algorithms is to divide those that aim to converge to the best result (to an optimal policy) and those that only track the payoff of different solutions (no convergence guarantees). An analysis of these two approaches in a stationary task found that in certain cases a tracking algorithm obtains better results than one that converges to the optimal policy~\citep{Sutton:2007ha}. This is especially important when dealing with non-stationary environments. When a change in the environment occurs a converging algorithm may take longer to overcome this issue~\citep{Claus:1998tb} and one that is only tracking will be able to adapt faster. This relates to the \emph{transient performance}, where usually algorithms are more concerned with the results of learning than with the ongoing process of learning~\citep{Sutton:2007ha}.

\item{Tolerated and induced non-stationarity.} There are different types of theoretical results in multiagent learning (e.g., convergence, optimality, non-exploitability; see Section~\ref{sec:theoreticalresults}). However, more general convergence results are needed and we propose two concepts that are worth analysing in MAS: \emph{tolerated non-stationarity}, this is, how much non-stationarity does an algorithm accepts without sacrificing optimality; and \emph{induced non-stationarity}, this is, how much non-stationarity an algorithm induces in the system. 

\end{itemize}

\vspace{1cm}
\paragraph{Line 4: Applications}

\begin{itemize}
\item{Negotiation and MAS.} 
As described in Section~\ref{sec:experimentaldomains}, negotiation is an interesting and real-world scenario to model multiagent interactions. However, generic negotiation using reinforcement learning seems an understudied subject with few works in the intersection (e.g.,~\citealp{Lazaric:2007ws}), as most research in this area seems to have focused on \qlearning for trading agents in competitive market places so far~\citep{Hsu:2001,Tesauro2002}. It would be interesting to employ a number of techniques mentioned in this survey (e.g., \citealp{Johanson:2007ts,Crandall:2011dt,Wunder:2014wu,HernandezLeal:2016tw}) in order to improve generic preference learning and strategy estimation in automated negotiation, both in bilateral and multilateral settings. \textcolor{black}{This remains an unsolved challenge in a non-stationary setting in which preference evolution can occur, for example with regard to risk tolerance or fairness attitudes~\citep{NegotiationChallenges}.}

%%%TODO: https://pdfs.semanticscholar.org/8b12/88f8d930daea8184e7528b53946791bc767c.pdf
\item{Deep RL and MAS.}
Deep learning~\citep{Bengio:2009kb} has shown outstanding results when combined with reinforcement learning~\citep{Mnih:2015jp,Silver:2016hl}. Even though most works assume a single-agent setting, problems with non-stationarity have already appeared, proposing extensions of existing algorithms that handle non-stationary environments in the deep learning setting. In particular, since deep learning approaches require large numbers of samples, common techniques such as experience replay have been adapted to handle non-stationarity~\citep{Foerster:2017ti,Castaneda:ui}. Moreover, deep multi-agent RL works are on the rise~\citep{He:2016up,Foerster:2016ud,Foerster:2017uq,Leibo:2017wi,Gupta:2017to,Tampuu:2017fc} with the obvious challenge of handling non-stationary environments (i.e., multiple learning agents). While these initial works have transferred a number of individual techniques to the deep setting, it remains an open challenge to provide a conceptual framework for deep multi-agent learning.
%\reviseq{Was not mentioned in the reviews but there's a ton of work here that we are not covering}

\end{itemize}
Above, we have presented relevant open problems with potential impact on the multiagent community. The next section presents the conclusions drawn from this survey.

% !TEX root = survey.tex

\section{Conclusions}
\label{sec:conclusions}

%\comment{(1) analyze a significant body of AI research and make it more accessible to a broader audience; (2) position existing research results in a broader context and explain their impact; (3) bring together previously unconnected lines of research in a way that fosters new research directions in these areas; (4) identify deficiencies or gaps in current knowledge that need to be addressed in future research.}

%\revise{we identified multi-agent interactions as key drivers of sustained innovation and, perhaps, increases in intelligence over the course of human evolution. As a consequence, a research program based only on replicating individual human cognitive abilities, e.g. attention, memory, planning, etc, is likely incomplete. It seems that intelligence researchers would do well to pay more attention to the ways in which multi-agent dynamics may structure both evolution and learning.\cite{autocurricula2019} }
Non-stationary environments in sequential decision making tasks have received attention from research in the domains of game theory, reinforcement learning and multi-armed bandits. This survey has reviewed a wide range of algorithms from these fields, and contributes a structure to think clearly about otherwise often implicit assumptions, characteristics and concepts related to the challenges of multiagent learning (see Section~\ref{sec:learningMAS}). \textcolor{black}{First, we proposed a new framework for reasoning about multiagent systems (see Section~\ref{sec:newFramework})}. Then, we identified several principled approaches that algorithms take to deal with non-stationarity:  ignore, forget, respond to target opponents, learn opponent models and theory of mind (see Section~\ref{sec:categories}). For each category we provide an illustrative example (see Section~\ref{sec:example}) and later we present an extensive list of state-of-the-art algorithms classified into these categories (see Section~\ref{sec:learningnonstationary}). Moreover, we identified the strengths and limitations of each category and provide guideline scenarios when they should be applied (see Section~\ref{sec:strenghtsandlimitations}).

We observed that most experimental results are formalised in terms of repeated games and stochastic games (see Section~\ref{sec:experimentaldomains}). Theoretical results are diverse and include: guarantees to learn optimal policies, non-exploitability guarantees and convergence to equilibria, to name a few (see Section~\ref{sec:theoreticalresults}).
Following the coherent review of the state of the art, this survey pinpoints the remaining open questions and presents them clustered into four open avenues for promising future research: diversity in opponents, dynamic interactions, learning objectives and applications (see Section~\ref{sec:openquestions}).

% Each paper should have a memorable closing sentence
While much progress has been achieved over the last decades, further fundamental research is required for the breakthrough guarantees and demonstration of algorithmic performance in non-stationary environments.
This survey seeks to facilitate this future work by highlighting current gaps in the literature and providing the guideline taxonomy to position future work within it.
  
%\item Many algorithms learn models of the agents in the environment which is used for two different purposes. The model can be used to track those for possible changes \citep{Jensen:2005ui,Marinescu:2015wt,DaSilva:2006uw,HernandezLeal:2013dq, HernandezLeal:2015tq,HernandezLeal:2016twa} and in most of the cases the model is used to derive an acting policy, however, there are some algorithms in which the opponent model is completely independent of the acting policy \citep{MohamedElidrisi:2012wm}. 

\acks{We would like to thank Frans Oliehoek and Daan Bloembergen for useful discussions and suggestions. This work is part of the Veni research programme with project number 639.021.751, which is financed by the Netherlands Organisation for Scientific Research (NWO).

%This research has received funding through the ERA-Net Smart Grids Plus project Grid-Friends, with support from the European Union's Horizon 2020 research and innovation programme.
}

\bibliographystyle{acm}
\bibliography{bibAll,librosDoctorado,extrabib}

\end{document}